\documentclass[12pt,a4paper]{article}  %,hidelinks,draft
\RequirePackage{ifpdf}
\usepackage{fixltx2e}

\def\thetitle{%On the 
The 
one-loop effective potential of the Wess-Zumino model
revisited}		\title{\thetitle} 
\def\theauthor{Simon J.\ Tyler \and Sergei M.\ Kuzenko}						\author{\theauthor}
\def\thedate{\today}												\date{\thedate}

\parskip=\medskipamount

\usepackage[mathscr]{euscript} 	% alternate script style
\usepackage{bbm} 
\usepackage{amsfonts,amsmath,amssymb} %,mathtools
\usepackage[square, comma, numbers, sort&compress]{natbib}	% remove me
\usepackage{moreverb}		% verbatimtab
\ifpdf
  \usepackage[pdftex]{graphicx}	
  \usepackage[dvipsnames,usenames,pdftex]{color}
  \usepackage[pdftex,bookmarks]{hyperref}	%,pagebackref
\hypersetup{bookmarksopen,bookmarksnumbered,
	pdfnewwindow=true,
	pdfauthor = {Simon J. Tyler, Sergei M. Kuzenko},
	pdftitle = {\thetitle},
	pdfsubject = {Physics},
	pdfkeywords = {supersymmetry, effective action},
	pdfcreator = {LaTeX with hyperref package},
	pdfproducer = {pdflatex}
}
\else
  \usepackage[dvips]{graphicx}
  \usepackage[dvipsnames,usenames,dvips]{color}
  \usepackage[dvips,bookmarks]{hyperref} %,pagebackref
\fi
\newcommand {\cD}{{\cal D}}

\newcommand {\cG}{{\cal G}}

\newcommand {\cN}{{\cal N}}

\newcommand {\cP}{{\cal P}}

\newcommand{\dsA}{{\mathbb A}}
\newcommand{\dsa}{{\mathbf A}} % Used to be bb a

\newcommand{\dsJ}{{\mathbb J}}

\newcommand{\dsL}{{\mathbb L}}

\def\ds1{\ensuremath{\mathbbm{1}}}
\def\a{\alpha}
\def\b{\beta}
\def\c{\chi}
\def\d{\delta}

\def\f{\phi}
\def\g{\gamma}
\def\G{\Gamma}

\renewcommand{\j}{\psi}

\def\l{\lambda}
\def\m{\mu}

\def\p{\pi}
\def\q{\theta}

\def\t{\tau}

\def\z{\zeta}
\def\D{\Delta}
\def\F{\Phi}
\def\J{\Psi}

\def\O{\Omega}

\def\S{\Sigma} 		% overwrites standard alias for \textsection / \mathsection

\def\X{\Xi}
\def\eps{\varepsilon}
\def\vf{\varphi}
\newcommand{\abar}{{\bar a}}
\newcommand{\bbar}{{\bar b}}

\newcommand{\Db}{{\bar D}}

\newcommand{\Fb}{{\bar\Phi}}
\newcommand{\fb}{{\bar\f}}
\newcommand{\Jb}{{\bar\Psi}}

\newcommand{\vfb}{{\bar\varphi}}
\newcommand{\qb}{{\bar\q}}
\newcommand{\scF}{\mathscr{F}}
\newcommand{\scG}{\mathscr{G}}

\newcommand{\scJ}{\mathscr{J}}

\newcommand{\rmd}{{\rm d}}
\newcommand{\rme}{{\rm e}}
\newcommand{\rmi}{{\rm i}}

\newcommand{\da}{{\dot{\alpha}}}
\newcommand{\db}{{\dot{\beta}}}

\newcommand{\ada}{{\alpha\dot\alpha}}

\newcommand{\daa}{{\dot\alpha\alpha}}

\newcommand{\bda}{{\beta\dot\alpha}}

\newcommand{\dba}{{\dot\beta\alpha}}

\newcommand{\be}{\begin{equation}}
\newcommand{\ee}{\end{equation}}
\newcommand{\bea}{\begin{eqnarray}}
\newcommand{\eea}{\end{eqnarray}}
\newcommand{\non}{\nonumber}
\newcommand{\bem}{\begin{pmatrix}}
\newcommand{\eem}{\end{pmatrix}}

\def\intx{\int\!\!{\rmd}^4x\,}

\def\intdk{\int\!\!\frac{\rmd^dk}{(2\p)^d}\,}
\def\intz{\int\!\!{\rmd}^8z\,}

\def\intc{\int\!\!{\rmd}^6z\,}
\def\intac{\int\!\!{\rmd}^6\bar z\,}
\newcommand{\dpm}{\bem \d_+ & 0 \\ 0 & \d_- \eem}	% (anti)Chiral Dirac delta

\newcommand{\pd}{\partial}
\newcommand{\ord}[1]{\ensuremath{{\rm O}\!\!\,\left(#1\right)}}
\newcommand{\acom}[2]{\ensuremath{\left\{ #1 , #2 \right\}}}
\newcommand{\com}[2]{\ensuremath{\left[ #1 , #2 \right]}}

\def\cc{{\rm c.c.\ }}

\newcommand{\BKY}{\ensuremath{\text{BKY}}}
\def\Tr{{\rm Tr}} 				% functional trace
\def\bTr{{\rm\bf Tr}} 			% functional trace incl chiral/antichiral matrix
\def\bDet{{\rm\bf Det}}
\def\const{{\rm const}}
 
\newcommand{\mub}{\bar{\mu}}		% \mub^2 = 4\pi\exp(-\gamma)\mu^2
\newcommand{\Li}{\mathrm{Li} }

\newcommand{\At}{\tilde{A}}
\newcommand{\Bt}{\tilde{B}}
\newcommand{\Ct}{\tilde{C}}
\newcommand{\JbJm}{\bem\Jb\\\J\eem}

\newcommand{\jj}{{m'}}  		% \J| = m + \l\f, the background field dependant mass in components

\begin{document}                        %%%%%%%%%%%%%%%%%%%%%%%%%%%%%%%%%%%%%%%%
%%%%%%%%%%%%%%%%%%%%%%%%%%%%%%%%%%%%%%%%%%%%%%%%%%%%%%%%%%%%%%%%%%%%%%%%%%%%%%%%
\begin{titlepage}

\begin{flushright} \thedate \end{flushright}

\vspace{5mm}

\begin{center}

{\large \bf  \thetitle}

{\large  
{Sergei M.\ Kuzenko}
%\footnote{sergei.kuzenko@uwa.edu.au}
and {Simon J.\ Tyler}\footnote{simon.tyler@graduate.uwa.edu.au}}

\vspace{5mm}

\footnotesize{
{\it School of Physics M013, The University of Western Australia\\
35 Stirling Highway, Crawley W.A. 6009, Australia}}
\end{center}

\vspace{5mm}

\pdfbookmark[1]{Abstract}{abstract_bookmark}
\begin{abstract}
\baselineskip=14pt

The full one-loop supersymmetric effective potential for the Wess-Zumino model 
is calculated using superfield techniques. 
This includes the K\"ahler potential and the auxiliary field potential,
of which the former was originally computed in 1993 while
the latter is derived for the first time. In the purely bosonic sector
our results match those of older component field calculations.

In light of prior contradictory results found in the literature, 
the calculation of the leading term in the auxiliary field potential
is approached in a variety of ways. 
Issues related to conditional convergence that occur during 
these calculations and their possible consequences are discussed.

\end{abstract}
\vspace{.5cm}

\begin{flushright} {\it  In memory of Professor Bruno Zumino} \end{flushright}
\vfill

\end{titlepage}

\newpage

%%%%%%%%%%%%%%%%%%%%%%%%%%%%%%%%%%%%%%%%%%%%%%%%%%%%%%%%%%%%%%%%%%%%%%%%%%%%%%%%
\section{Introduction}
\setcounter{equation}{0}
\label{sec:Intro}

%%%%%%%%%%%%%%%%%%%%%%%%%%%%%%%%%%%%%%%%%%%%%%%%%%%%%%%%%%%%%%%%%%%%%%%%%%%%%%%%

The Wess-Zumino (WZ) model was proposed forty years ago
\cite{Wess1974a,Wess1974b}.
It is the third oldest supersymmetric field theory in four dimensions.\footnote{Only
the  supersymmetric massive QED of Golfand and Likhtman 
\cite{Golfand} and the Goldstino model of Volkov and Akulov \cite{VA}
appeared before the WZ model.} 
It is  the first  {\it off-shell} and {\it renormalised} supersymmetric  
field theory ever constructed. 
As such, it has long acted as 
both a testbed and a teaching tool for supersymmetry.

We recall that the classical action for the WZ model 
is\footnote	{We follow the conventions and notation of \cite{BK}.}  
\begin{align} \label{eq:WZ_Action}
	S[\F,\Fb] &= \intz \Fb\F
		+ \intc\cP(\F) + \intac\bar\cP(\Fb) \,,
\end{align}
where $\cP (\F) $ denotes the superpotential 		
\begin{align}
	\cP(\F) &= \frac m2 \F^2 + \frac \l6 \F^3 \,,
\end{align}
with $m$ and $\l$ constant parameters. The dynamical variables are 
a  chiral scalar superfield $\F$, $\Db_\da\F=0$, 
and its complex conjugate $\bar \F$.
The superpotential, $\cP(\F)$, must be at most cubic 
for the model to be renormalisable. 
The mass parameter $m$ can always be chosen to be real and non-negative. 
The coupling constant $\l$ is complex in general. 
In the massless case, $m=0$, the WZ action is superconformal 
\cite{Wess1974a,Wess1974b}.

This paper primarily focuses on the one-loop quantum corrections 
to the  effective potential of the WZ model, in particular, 
on the auxiliary field potential defined below. 
In superspace, the full effective action of the WZ model has the generic form
\cite{Buchbinder1993, Buchbinder1994a, BK}
\begin{align}\label{eq:GenEffAct}
	\G[\F,\Fb]&=\intz\dsL(\F,D_A\F,
	\dots,\Fb,D_A\Fb,
	\dots)
	+\left(\intc\dsL_c(\F)+\cc\right)\,,
\end{align}
where $\dsL=\Fb\F+\ord\hbar$ is the effective superspace Lagrangian, 
$\dsL_c=\cP+\ord\hbar$ is the effective superpotential,\footnote{When
  all fields are massive, the (chiral) superpotential does not receive any
	quantum corrections, this was one of the earliest 
	supersymmetric nonrenormalization theorems 
	\cite{Wess1974b,Iliopoulos1974,Ferrara1974,Delbourgo1975,
	West1976,Weinberg1976,Grisaru1979}. 
	When there are massless fields present, 
	finite corrections to the superpotential can exist \cite{GGRS1983,West}.
	In the massless WZ model, 
	the first quantum correction to the superpotential  occurs at two loops. 
	It was originally calculated in components by 
	Jack {\it et al.}
		in \cite{West1991} 
	and then using superfield methods by 
	Buchbinder
		{\it et al.} in \cite{Buchbinder1994}.
	No chiral superpotential is generated at the quantum level  
	if one starts from the effective action 
	for the massive WZ model and then considers a massless limit.}
with $\cP (\F)$  the classical superpotential.
 In the first term 
 in the right-hand side of \eqref{eq:GenEffAct}, 
 $D_A$ denotes the superspace covariant derivatives,
  $D_A = ( \pd_a , D_\a, \Db^\da)$.
For fields constant in spacetime,  $\pd_a\F=\pd_a\Fb=0$,
the effective superspace Lagrangian decomposes into
\begin{align} 
\label{defn:WZ-KahlerPot}
	\dsL\big|_{\pd_a\F=\pd_a\Fb=0}&=K(\F,\Fb)
	+ \mathbb  F(\F,D_\a\F,D^2\F,\Fb,\Db_\da\Fb,\Db^2\Fb)\,,
\intertext{where \vspace{-1.5em}}
	K&=\Fb\F+\sum_{n=1}^{\infty}\hbar^n K^{(n)}\,, \\ 
\intertext{is the effective K\"ahler potential,  and}
\label{defn:WZ-AuxPot}
	\mathbb  F&=\sum_{n=1}^{\infty}\hbar^n \mathbb  F^{(n)}\,,
	\quad \mathbb  F\big|_{D_\a\F=\Db_\da\bar\F=0}=0\,,
\end{align}
is called the effective auxiliary field potential (EAFP). 
The name for $\mathbb  F$ is appropriate since, when reduced to components
in a constant background, the EAFP
is of at least third order in the auxiliary fields
\cite{Buchbinder1994a,BK}. 
Modulo total derivatives and terms proportional to $\pd_a \F$ and $\pd_a \bar \F$, 
the EAFP
can always be reduced to the form
\begin{align}\label{eqn:Gen_Aux_Pot}
	\mathbb  F &= (D^\a\F)(D_\a\F)(\Db_\da\Fb)(\Db^\da\Fb)
				{\mathbb  G}
		(\F,D^2\F,\Fb,\Db^2\Fb)\, .
\end{align}
This shows that its leading term must have at least four spinor derivatives. 
The supersymmetric effective potential is determined by $K$, $\mathbb F$ and 
$\dsL_c$.

Using the component formulation of the model, 
the one-loop correction to the effective potential of  the WZ model
was calculated
in the year following the model's proposal \cite{Fujikawa1975}.
Subsequently this was extended by many authors to include 
more general models, higher loops and superspace based calculations  
\cite{O'Raifeartaigh1976a,Huq1977,Amati1982,
Grisaru1983a,Miller1983a,Miller1983,Fogleman1983}.
However, all superspace calculations used a background chiral superfield that
did not include a spinor component,
\begin{align}\label{eq:spurion_background}
	\F( \q) = \f + \q^2 F\,, \qquad \pd_a\f = \pd_aF = 0\ , 
\end{align}
thus breaking explicit supersymmetry. 
In all of these papers, the effective potential was always computed as a function 
of the scalars $\f$, $F$ and their conjugates, and never  as a superspace Lagrangian 
of the form \eqref{defn:WZ-KahlerPot}.  The point is that, within the standard 
supergraph technique \cite{Grisaru1979}, the problem of computing the quantum corrections 
to $\mathbb F$ is analogous to that of computing quantum corrections with derivatives of 
fields in ordinary scalar field theories. Actually, in order to determine 
$\mathbb  F$, one has to evaluate quantum corrections with an {\it arbitrarily} large 
number of spinor covariant derivatives, which appears to be a daunting task. 

The first manifestly  supersymmetric calculation of the effective potential
of the WZ model was not until 1993 \cite{Buchbinder1993, Buchbinder1994a}.
These papers developed a superfield heat kernel technique to compute quantum 
corrections to the effective potential of the WZ model. The one-loop 
effective action was expressed in terms of the Green function 
for a real scalar superfield in the presence of a background chiral scalar $\F$.
The heat kernel corresponding to this superpropagator was computed exactly 
in the case when $\F$ satisfies
the supersymmetric constraint $\pd_a\F=0$ and has the explicit form 
\begin{align}
	\F(\q) = \f + \q^\b \j_\b + 
	\q^2 F\,, \qquad \pd_a\f = \pd_aF = 0\ ,
	\quad \pd_a \j_\b=0	\ . 
\label{1.7}
\end{align}
Due to the presence of the spinor $\j_\b$ 
in the background superfield, 
the heat kernel
derived in \cite{Buchbinder1993, Buchbinder1994a} 
is significantly more complicated than that which occurs when using
the non-supersymmetric background \eqref{eq:spurion_background}.

The heat kernel
derived in \cite{Buchbinder1993, Buchbinder1994a} suffices
to compute the one-loop supersymmetric effective potential  
\eqref{defn:WZ-KahlerPot} exactly, which will be done in this paper.
However, explicit calculations were given in 
\cite{Buchbinder1993, Buchbinder1994a} 
only for two special structures: 
the K\"ahler and the leading four-derivative contribution to  the EAFP.
The one-loop K\"ahler potential was found to be 
\bea
K^{(1)} = -\frac{1}{2(4\p)^2} |\cP'' (\F)|^2 \left( 
\ln 
\frac{ |\cP'' (\F)|^2 }{ \m^2} 
- 2\right) ~,\qquad 
 \cP'' (\F) = m + \l \F~,
 \label{1.10}
\eea
with $\m$ the renormalisation scale.
The four-derivative  correction to $\mathbb  F$ was found to be
\bea
{\mathbb  F}^{(1)}_{\text{4-deriv}} 
	= \z\,  \frac{ |\l | ^4}{(4\p)^2 }		
	\frac{(D^\a\F )( D_\a\F)(\Db_\da\Fb)(\Db^\da\Fb) }{ | \cP'' (\F) |^4} \,, 
\label{WZ-4Deriv-Zeta}
\eea
for some numerical coefficient $\z$. This coefficient was not evaluated
explicitly in \cite{Buchbinder1993, Buchbinder1994a}, 
but an integral representation for $\z$ was given.  
In what follows, it will be referred  to  
as  $\z_{\BKY}$. Since the four-derivative correction \eqref{WZ-4Deriv-Zeta}
is UV finite, no regularisation was used 
in \cite{Buchbinder1993, Buchbinder1994a} for its evaluation. 

The one-loop K\"ahler potential, $K^{(1)}$, and the leading contribution to the 
EAFP, ${\mathbb  F}^{(1)}_{\text{4-deriv}} $,
were subsequently recalculated using 
supergraphs
\cite{Pickering1996}
and a covariant superfield derivative expansion \cite{Pletnev1999}. 
Both of these methods are equivalent in spirit to the expansion 
described in section \ref{sec:DirectExpn} 
and would be cumbersome to take to higher orders.

As pointed out in \cite{Buchbinder1993, Buchbinder1994a}, 
the K\"ahler potential is much easier to calculate 
than the full supersymmetric effective potential.
This is because during the calculation it suffices to use the condition 
$D_\a\F=0$ for the background field, leading to much simpler propagators.
This allows for calculations in more general models and at higher loops
\cite{Grisaru1996, DeWit1996, Nibbelink2005}.

Despite the long history described above, 
many interesting aspects remain to be explored in 
the calculation of the one-loop supersymmetric effective potential of the WZ model.
In this paper we further examine the issues arising in 
the superfield calculation of the one-loop EAFP. 
This was motivated by the observation that the direct 
evaluation of the integral $\z_{\BKY}$ given in
\cite{Buchbinder1993, Buchbinder1994a} did not match 
the value of $\z$ found 
in the later papers \cite{Pickering1996, Pletnev1999}, 
which used dimensional regularisation. 
We resolve this issue by repeating the earlier calculations and 
demonstrating that the result is ambiguous due to conditionally convergent integrals.
However, using dimensional regularisation fixes the result and yields a coefficient that matches the corresponding term in the earlier component results.
We then proceed to use our techniques to present the first superfield calculation of the full one-loop EAFP and compare it to the component results.

Before turning to the computational aspects of this paper, 
we would like to discuss the functional 
form of the four-derivative quantum correction \eqref{WZ-4Deriv-Zeta}.
In the case of the massless WZ model, the expression on the right 
of \eqref{WZ-4Deriv-Zeta}
becomes $\l$-independent and  
proportional to 
\bea
 \frac{(D^\a\F )( D_\a\F)(\Db_\da\Fb)(\Db^\da\Fb) }{ (\F \bar \F )^2} 
= (D^\a \ln \F )( D_\a \ln \F)(\Db_\da \ln \Fb)(\Db^\da \ln \Fb) ~.
\eea
For the massless WZ model,
it is more advantageous to define the four-derivative quantum 
correction in a somewhat different form as follows
\begin{subequations}\label{1.14}
\bea
\widetilde{\mathbb  F}^{(1)}_{\text{4-deriv}} 
	&=&  \frac{ \z}{(4\p)^2 } \X ~, \label{1.14a}
	\eea
where we have introduced
\bea
	\X&:=&
\Big[ (D^\a \ln \F )( D_\a \ln \F)+ (D^2 \ln \F) \Big]
\Big[(\Db_\da \ln \Fb)(\Db^\da \ln \Fb) 
+
(\bar D^2 \ln \bar \F) \Big] \non \\
&=& \frac{(D^2 \F)(\bar D^2 \bar \F)}{\F \bar \F}
~.~~~~
\label{1.13b}
\eea
\end{subequations}
It  holds that 
$ \intz  \widetilde{\mathbb  F}^{(1)}_{\text{4-deriv}} 
\approx  \intz  {\mathbb  F}^{(1)}_{\text{4-deriv}} $
modulo the terms proportional to vector derivatives of $ \F$ and $\bar \F$.
The main advantage of the new definition \eqref{1.14a} is that 
$$(D^\a \ln \F )( D_\a \ln \F)+ (D^2 \ln \F) = \frac{D^2 \F}{\F} $$
is a (conformal) primary  superfield
such that the functional $ \intz \X$
is invariant under the $\cN=1$ superconformal transformations
(see \cite{BK} for a review on  $\cN=1$ superconformal field theories). 
We recall that the massless WZ model is superconformal 
at the classical level \cite{Wess1974a,Wess1974b}.
Of course, the superconformal symmetry is anomalous in the 
quantum theory. However, it is the effective K\"ahler potential 
which encodes the information about the superconformal anomaly. 
It is quite natural to define the EAFP to be superconformal.

In the massless case, the entire one-loop EAFP may be chosen to be a primary superfield 
of the form
\bea
\widetilde{\mathbb  F}^{(1)}_{\rm massless} = \X \,{\mathbb  H} \left(\frac{\X}{\F \bar \F}\right)~,
\label{1.15}
\eea
for some real function $\mathbb  H (x) $ such that $\mathbb  H (0) = \z /(4\p)^2$.
It may be seen that $ \intz  \widetilde{\mathbb  F}^{(1)} $ is invariant under the superconformal transformations. In accordance with \cite{Buchbinder1994a}, 
the one-loop effective action of the massless WZ model is invariant 
under  phase transformations $\F \to {\rm e}^{\rm i \t} \F$,
with $\t$ a constant parameter. 

The structure of this paper is as follows. 
In section \ref{sec:Quant} we examine the quantisation of the WZ model and the structure of its one-loop effective action.
In section \ref{sec:DirectExpn} we use a brute force approach that emulates the diagrammatics of \cite{Pickering1996} 
to calculate the one-loop K\"ahler potential and leading term to the EAFP \eqref{WZ-4Deriv-Zeta}.
In section \ref{sec:AuxEffPot} we use the heat kernel of appendix \ref{sec:WZProp} to calculate the one-loop
K\"ahler potential as well as the leading correction and the full expression for the 
EAFP.
The component results for the effective potential and their comparisons to the superfield results are given in section \ref{sec:Components}.
In the last section of this paper summarises the results and looks at the further work that could be done.
The paper contains one \hyperref[sec:WZProp]{appendix} 
that repeats the calculation of \cite{Buchbinder1993, Buchbinder1994a}
to find the heat kernel for the WZ model. 
The result is put into the simplest form possible and the K\"ahler limit is investigated.

Most of the original results given in this work first appeared in the PhD thesis \cite{TylerPhD2013} and many more details can be found
in that text and the accompanying auxiliary {\it Mathematica} files.
Section 2 and appendix A are comprised of review material from 
 \cite{BK,Buchbinder1993, Buchbinder1994a}.

%%%%%%%%%%%%%%%%%%%%%%%%%%%%%%%%%%%%%%%%%%%%%%%%%%%%%%%%%%%%%%%%%%%%%%%%%%%%%%%%

\section{Quantization}
\setcounter{equation}{0}
\label{sec:Quant}

%%%%%%%%%%%%%%%%%%%%%%%%%%%%%%%%%%%%%%%%%%%%%%%%%%%%%%%%%%%%%%%%%%%%%%%%%%%%%%%%

The functional integral representation for the effective action \eqref{eq:GenEffAct}
is \cite{BK,Buchbinder1994a}
\begin{align} \label{WZEffectiveAction}
	\rme^{\frac\rmi\hbar\tilde\G[\F,\Fb]}
	= \cN \int \!\! \cD\vf\cD\bar\vf \; \rme^{\frac\rmi\hbar S^{(\J)}[\vf,\bar\vf]
	+\rmi\hbar^{1/2} S_{\rm int}[\vf,\bar\vf]
	-\rmi\hbar^{-1/2}\left(\vf\cdot\frac{\d\tilde\G}{\d\F}
		+\bar\vf\cdot\frac{\d\tilde\G}{\d\Fb}\right) } \,, 
\end{align}
where $\tilde\G[\F,\Fb]=\G[\F,\Fb]-S[\F,\Fb]$ and we have introduced
the background chiral scalar 
\begin{align} 
\label{WZDefnOfPsi}
    \J	:=\cP^{''}(\F) =m+\l\F
\end{align}
and the action 
\begin{subequations}
\begin{align}
\label{WZFreeAction}
	S^{(\J)}[\vf,\bar\vf]&=\intz \bar\vf\vf 
		+ \frac12\left(\intc\J\vf^2+\cc \right) \,, \\
\label{WZIntAction}
	S_{\rm int}[\vf,\bar\vf]&=\frac\l6\intc\vf^3 +\cc
\end{align}
\end{subequations}
From the above, it is clear that the effective action depends on 
$\F$ only through the combination $\J$.
The only interaction terms in the theory are the cubic vertices of 
\eqref{WZIntAction}, however, these are not needed 
in the one-loop calculations of this paper.

\subsection{Propagators of the WZ model}
To find the propagators for the model, we note that
the Hessian for the free action \eqref{WZFreeAction} is
defined by 
\begin{align}\label{eqn:WZ_Hessian}
	S^{(\J)}[\vf,\bar\vf] &= 
		\frac12(\vf,\vfb)\cdot H^{(\J)} \cdot \bem\vf\\\vfb\eem \,,\quad
	H^{(\J)} = \bem \J&-\frac14\Db^2\\-\frac14 D^2&\Jb\eem\dpm .
\end{align}
Where the functional inner product ``$\cdot$'' is a matrix product 
as well as the integration over the appropriate superspaces,
we suppress the superspace coordinates and use the 
chiral/antichiral delta function matrix
\begin{align}\label{def:dpm}
	\dpm = -\frac14\bem \Db^2 && 0\\0 && D^2 \eem\d^8\ .
\end{align}
We can invert the Hessian by writing 
\(G^{(\J)}=-H^{(0)}\cdot(H^{(\J)}\cdot H^{(0)})^{-1}\)
and using the block matrix inverse formula 
to get
\begin{align*}
	G^{(\J)} =
	\bem
	\frac1{16}\Db^2\frac{1}{\Box-\frac1{16}\Jb D^2\frac1\Box\J\Db^2} 
		\Jb\frac1\Box D^2 \!\!\! && \!\!\!
	\frac14\Db^2\frac{1}{\Box-\frac1{16}\Jb D^2\frac1\Box\J\Db^2} \\
	\frac14 D^2\frac{1}{\Box-\frac1{16}\J \Db^2\frac1\Box\Jb D^2} \!\!\!&&\!\!\!
	\frac1{16} D^2\frac{1}{\Box-\frac1{16}\J \Db^2\frac1\Box\Jb D^2}
		\J \frac1\Box \Db^2
	\eem \dpm \,,
\end{align*}
where, for the rest of this paper we use the convention that 
\emph{all} derivatives act on \emph{all} terms to the right 
unless bracketed or otherwise indicated.
After using equation \eqref{def:dpm}, 
expanding the inverses as a geometric series and performing some $D$-algebra,
the Green function becomes
\begin{align*}
	G^{(\J)} 
	&= \frac1{16}\sum_{n=0}^\infty\bem
	\Db^2(-\frac1\Box\frac{\Jb D^2}{-4}-\frac1\Box\frac{\J\Db^2}{-4})^n\Db^2  && 
	\Db^2(-\frac1\Box\frac{\Jb D^2}{-4}-\frac1\Box\frac{\J\Db^2}{-4})^n D^2	\\ 
	D^2(-\frac1\Box\frac{\Jb D^2}{-4}-\frac1\Box\frac{\J\Db^2}{-4})^n \Db^2   &&
	D^2(-\frac1\Box\frac{\Jb D^2}{-4}-\frac1\Box\frac{\J\Db^2}{-4})^n D^2
	\eem \frac{-1}\Box\d^8 \ .
\end{align*}
Resumming the above series recovers the result of \cite{Buchbinder1994a} 
and we see that the Green function can be written in the form
\begin{align} \label{WZ-Green's Function}
		G^{(\J)}(z,z')
	&= \bem  G^{(\J)}_{++}(z,z') & G^{(\J)}_{+-}(z,z') \\ 
			 G^{(\J)}_{-+}(z,z') & G^{(\J)}_{--}(z,z') \eem
	 = \frac1{16}\bem \Db^2\Db'^2 & \Db^2 D'^2 \\ D^2\Db'^2 & D^2D'^2\eem
	G_V^{(\J)}(z,z')\,,
\end{align}
where the auxiliary Green function $G_V^{(\J)}$, introduced in 
 \cite{Buchbinder1993, Buchbinder1994a},  satisfies the equation
\begin{align} \label{WZ-AuxGreen’s Function}
	\D G_V^{(\J)}(z,z') = -\d^8(z,z')\,, 
	\quad\text{with}\quad
	\D = \Box-\frac14\J\Db^2-\frac14\Jb D^2 \ .
\end{align}
This auxiliary propagator can be understood in terms of its 
heat kernel representation 
\begin{subequations} \label{2.8}
\begin{gather}
	G_V^{(\J)}(z,z') = \rmi \int_0^\infty U_V^{(\J)}(z,z'|s) \rmd s \,, \\
	(\pd_s - \rmi \D) U_V^{(\J)}(z,z'|s) = 0 \,, \quad
	U_V^{(\J)}(z,z'|0) = \d^8(z,z') \ .
\end{gather}
\end{subequations}

In the constant background $\pd_a\J=\pd_a\Jb=0$ that is the main focus of this paper, 
the heat kernel factorises to
\begin{align}\label{factored_U}
	U_V^{(\J)}(z,z'|s) = \O(s) U_V^{(0)}(z,z'|s) \,,
\end{align}
where $U_V^{(0)}(z,z'|s)=\d^4(\q-\q') U(x,x'|s)$ is the bosonic heat kernel
\begin{align}\label{bosonic_U}
	U(x,x'|s) = \frac{-\rmi}{(4\p s)^2} \rme^{\frac\rmi{4s}(x-x')^2}\ .
\end{align}
The operator $\O(s)=\rme^{-\frac{\rmi s}{4}(\J\Db^2+\Jb D^2)}$ can be expanded in 
powers of  spinor derivatives
\begin{align}
\begin{aligned}
	\O(s) 
	&= \frac1{16} A D^2 \Db^2 + \frac1{16}\tilde{A}\Db^2 D^2 
	+ \frac18 B^\a D_\a \Db^2 + \frac18 \tilde{B}_\da \Db^\da D^2
	+ \frac14 C D^2 + \frac14 \tilde{C} \Db^2 + 1 \,, \!\!\!
\end{aligned}
\end{align}
where $A,\dots,\tilde{C}$ are functions of $\pd_a$, $\J$, $D_\a\J$, $(D^2\J)$ and their complex conjugates.
The expressions for these functions are derived in detail in appendix \ref{sec:WZProp}.

\subsection{One-loop effective action}
The one-loop effective action can be written as the functional determinant 
obtained by turning off the interactions \eqref{WZIntAction} and performing 
the remaining Gaussian functional integral \eqref{WZEffectiveAction} to get
\begin{align} \label{1loopfunctional}
	\G^{(1)} = \frac\rmi2\log\bDet (H^{(\J)}/H^{(0)}) 
	=\frac\rmi2 \bTr\log (H^{(\J)}/H^{(0)}) \ ,
\end{align}
where the functional determinant and trace follow from the inner product defined in \eqref{eqn:WZ_Hessian}.
In particular, the full functional trace decomposes into a trace over the chiral and antichiral subspaces.
\begin{align}\label{defn:bTr}
	\bTr\bem A_{++} & A_{+-} \\ A_{-+} & A_{--} \eem
	= \Tr_+ A_{++} + \Tr_- A_{--}
\end{align}
The argument of the $\log$ in \eqref{1loopfunctional} is equivalent to
\begin{align}
	(H^{(0)})^{-1}\cdot H^{(\J)}
	=\left(1+\frac1\Box
		\bem 0&-\frac14 \Db^2\Jb\\-\frac14  D^2\J&0 \eem\right)\dpm \,,
\end{align}
and since only the diagonal terms survive the trace \eqref{defn:bTr}, we obtain
\begin{align} \label{WZ-1Loop-EffPot1}
	\G^{(1)}
	&=\frac\rmi4\Tr_+\log\left(1
		-\frac{\Db^2}{4\Box}\Jb\frac{D^2}{4\Box}\J\right) + \cc
\end{align}
By using both the the cyclicity of the functional trace 
and the fact that the trace over chiral superspace is equivalent to 
the chiral projection%
\footnote{
The $\cN=1$ superspace projection operators \cite{SaS,Sokatchev} are
\begin{subequations}\label{defn:SuSyProjOps}\begin{gather} 
	P_+ = \frac{\Db^2D^2}{16\Box}\,,\quad
	P_- = \frac{D^2\Db^2}{16\Box}\,,\quad
	P_0 = \frac{D^\a\Db^2D_\a}{-8\Box}
		= \frac{\Db_\da D^2 \Db^\da}{-8\Box}\,,\\
	P_iP_j=\d_{ij}\,,\quad P_0+P_++P_-=1\ .
\end{gather}\end{subequations}
} 
of the trace over full superspace $\Tr_+F_{++}=\Tr(F_{++}P_+)$,
we obtain two useful forms for the one-loop effective action
\begin{subequations}\label{WZ-1Loop-FunctionalForms}
\begin{align}
 \G^{(1)}
\label{WZ-1Loop-DirectEffPot}
 &=\frac\rmi4\Tr\sum_{n=1}^\infty\frac{-1}{n}\left(
 (P_+\Jb\frac1\Box\J)^nP_+ +\cc \right) \\ 
 \label{WZ-1Loop-HKEffPot}
 &=\frac\rmi2\Tr\sum_{n=1}^\infty\frac{-1}{n}\left(\frac{1}{\Box}
 \Jb\frac{D^2}{4}+\frac{1}{\Box}\J\frac{\Db^2}4\right)^n
 =\frac\rmi2\Tr\log\Big(\frac\D\Box\Big) \ .
\end{align}
\end{subequations}
The first form lends itself to a direct expansion of one-loop effective
potential performed in section \ref{sec:DirectExpn},
which is similar to the graphical expansion undertaken in \cite{Pickering1996}. 
The second form, 
which can also be derived starting from \eqref{WZ-Green's Function},
is used for the heat kernel based calculations of 
\cite{Buchbinder1993, Buchbinder1994a} and section \ref{sec:AuxEffPot}.

\pagebreak[2]
All of the above expressions hold in an arbitrary background; 
however, for the rest of this paper we will primarily focus on
the effective potential calculations in a constant background field $\pd_a\J=\pd_a\Jb=0$.
We find it convenient to use the following notation for 
the various combinations of derivatives of the background fields
\begin{align} \label{def:WZPropShortHand}
		&a=(D^\a\J)(D_\a\J)\,,\quad  \abar=(\Db_\da\Jb)(\Db^\da\Jb)\,,\quad
	b=(D^2\J)\,,\quad  \bbar=(\Db^2\Jb)\,, \non\\
	&\qquad\qquad u^2=\Jb\J\Box\,,\quad
	\scF^2=\bbar b/64\,,\quad \scG^2=u^2+\scF^2 \, .
\end{align}
For more details, see appendix \ref{sec:WZProp}.

%%%%%%%%%%%%%%%%%%%%%%%%%%%%%%%%%%%%%%%%%%%%%%%%%%%%%%%%%%%%%%%%%%%%%%%%%%%%%%%%

\section{Direct expansion of the one-loop effective action}
\setcounter{equation}{0}
\label{sec:DirectExpn}

%%%%%%%%%%%%%%%%%%%%%%%%%%%%%%%%%%%%%%%%%%%%%%%%%%%%%%%%%%%%%%%%%%%%%%%%%%%%%%%%

In this section we expand the expression for the one-loop effective action
\eqref{WZ-1Loop-DirectEffPot} and only keep up to the 4-derivative terms.
From \eqref{WZ-1Loop-DirectEffPot}, we see that we need to examine the term
\begin{align} \label{defn:Tn}
	T_n 
	:=
	(P_+\Jb\frac1\Box\J)^nP_+\,,
\end{align}
and its complex conjugate,
remembering that all derivatives, unless otherwise indicated, 
act on all terms to the right.
Since we're in the effective potential approximation 
we can commute all of the $\Box^{-1}$ terms to the left.  

We're interested in only the K\"ahler potential and the leading term in the auxiliary potential.
To calculate the K\"ahler potential, we can commute all of the derivatives 
and therefore all of the projection operators through the background fields to get
\begin{align} \label{eqn:Tn_Kahler}
	T_n \approx \Box^{-n}(\Jb\J)^nP_+\,.
\end{align}
To find the first term in the auxiliary potential,
we want to let a total of four Grassmann derivatives hit the backgrounds fields.
So, most of the chiral projectors will go straight through to the right;
however, there must be a first (from the right) 
chiral projector that hits a field, so we will need to sum over all possibilities:
\begin{align} \label{eqn:Tn_Aux1}
	T_n
	&= \Box^{-n}\sum_{j=0}^{n-1}(P_+\Jb\J)^{n-j-1}
	 \frac{\Db^2D^2}{16\Box}(\Jb\J)^{j+1}P_+\ .
\end{align}

%%%%%%%%%%%%%%%%%%%%%%%%%%%%%%%%%%%%%%%%%%%%%%%%%%%%%%%%%%%%%%%%%%%%%%%%%%%%%%%%

\subsection{K\"ahler potential}

${}$From \eqref{WZ-1Loop-DirectEffPot} and \eqref{eqn:Tn_Kahler} 
we see that we can resum the one-loop effective action in the K\"ahler approximation to get
\begin{align}
	\G^{(1)} &= \intz K^{(1)} 
	= \frac{\rmi}{4}\Tr\left(\log\left(1-\frac{\Jb\J}{\Box}\right)(P_++P_-)\right) \ .
\end{align}
Evaluating the trace by moving to momentum space with dimensional regularisation gives
the K\"ahler potential as 
\begin{align}
	K^{(1)} = -\frac{\rmi}{2}\mu^{2\eps} \intdk \frac1{k^2} \log\left(1+\frac{\Jb\J}{k^2}\right) \,,
\end{align}
where $d = 4 - 2\eps$ parametrises the dimensional regularisation scheme 
and $\m$ is the minimal subtraction renormalisation mass scale.
The integral can be performed to obtain
\begin{align} \label{eqn:1loopKahler}
\begin{aligned} 
	K^{(1)} &= \frac{1}{2} \scJ(\Jb\J) \\
	&= \frac{\m^{2\eps}}{2(4\pi)^{2-\eps}}\frac{\G(\eps)}{(1-\eps)^2}(\Jb\J)^{1-\eps} 
	= \frac{\Jb\J}{2(4\p)^2}\left(\frac1\eps + 2 - \log\frac{\Jb\J}{\mub^2} + \ord\eps\right)\,,
\end{aligned}
\end{align}
where $\mub^2=4\pi\rme^{-\g}\m^2$ is the modified minimal subtraction mass scale.
This result agrees with \eqref{1.10}.

%%%%%%%%%%%%%%%%%%%%%%%%%%%%%%%%%%%%%%%%%%%%%%%%%%%%%%%%%%%%%%%%%%%%%%%%%%%%%%%%

\subsection{Four-derivative term}

The four derivatives in the first projection operator to hit a field in \eqref{eqn:Tn_Aux1}
can hit the fields in many different ways. 
Summing over the possibilities we get the terms
\begin{align} T_n
	&= \Box^{-n}\sum_{j=0}^{n-1}(P_+\Jb\J)^{n-j-1}
	\Bigg( \com{\Db^2D^2}{(\Jb\J)^{j+1}}\frac{1}{16\Box} \non \\ \non 
	&\quad+ \com{\Db^2D^\a}{(\Jb\J)^{j+1}}\frac{D_\a}{8\Box}
	+\com{\Db^2}{(\Jb\J)^{j+1}}\frac{D^2}{16\Box} \\ \non
	&\quad+\com{\Db^\da D^\a}{(\Jb\J)^{j+1}}\frac{\Db_\da D_\a}{4\Box}
	+\com{\Db_\da}{(\Jb\J)^{j+1}}\frac{\Db^\da D^2}{8\Box} \Bigg)P_+  \\
	&:=
	T^{(1)}_n + T^{(2)}_n + T^{(3)}_n + T^{(4)}_n + T^{(5)}_n \ .
\end{align}
We will evaluate each term, $T_n^{(1,\dots,5)}$, in sequence.  
Note that for $n=1$, only the first term exists, 
but it is a total derivative and can thus be ignored.  

\subsubsection*{Evaluation of \texorpdfstring{$T^{(1)}_n$}{T^1_n}}
Since all four derivatives come from a single $P_+$, the rest of the 
projection operators commute through to the right,
\begin{align*} 
	T_n^{(1)} &= \frac{1}{\Box^n}\sum_{j=0}^{n-1}\frac{(\J\Jb)^{n-j-1}}{16\Box}
			\com{\Db^2 D^2}{(\J\Jb)^{j+1}}P_+ \\
	&= \frac{(\J\Jb)^{n-2}}{16\Box^{n+1}}\sum_{j=0}^{n-1} (j+1)^2
			(\Jb\bbar + j\abar)(\J b + j a) P_+ \ .
\end{align*}
Performing the simple sums of polynomials,
we find
\begin{align*} 
	T_n^{(1)} &= \frac{(\J\Jb)^{n-2}}{16\Box^{n+1}} n (n+1) \Big(
		  \abar a\frac{(n-1)(3n^2-2)}{15} \\
		&\qquad + (\abar\J b+a\Jb\bbar)\frac{(n-1)(3n+2)}{12}  
			+ \Jb\bbar\J b\frac{2n+1}{6}	 \Big)P_+ \\
		&= \abar a \frac{(\J\Jb)^{n-2}}{16\Box^{n+1}} \frac{n(n^4-1)}{30} P_+ 
			+ \text{surface terms} \ .
\end{align*}

\subsubsection*{Evaluation of \texorpdfstring{$T^{(2)}_n$}{T^2_n}}
The first projection operator provides three derivatives to give
\begin{align*} 
	T_n^{(2)} &= \frac{1}{8\Box^{n+1}}\sum_{j=0}^{n-1}(j+1)^2
		(P_+\Jb\J)^{n-j-1}\Jb^{j-1}(\Jb\bbar+j\abar)\J^j(D^\a\J)D_\a P_+ \ .
\end{align*}
Since $P_+D_\a P_+=0$, 
the final $D_\b$ to hit a field must come from the next projector on the right. 
This yields
\begin{align*} 
	T_n^{(2)} &= -\frac{(\Jb\J)^{n-2}}{8\Box^{n+1}}\sum_{j=0}^{n-1}(j+1)^2
		(\Jb\bbar+j\abar)(\J b + (j+1) a) P_+ \\
		&= \frac{-(\Jb\J)^{n-2}}{8\Box^{n+1}}
		\frac{n(n+1)}{60} \Big(\bar{a}a (n-1)\big(12 n^2+15 n+2\big)
			+5\bar{a}\J b (n-1)(3n+2) \\
		&\qquad + 15 a\Jb\bar{b} n(n+1) + 10\Jb\bar{b}\J b (2n+1) \Big)P_+\\
		&= -\abar a \frac{(\Jb\J)^{n-2}}{8\Box^{n+1}}
		\frac{(n-2)(n-1)n(n+1)(2n-1)}{60}P_+ + \text{surface terms} \ .
\end{align*}

\subsubsection*{Evaluation of \texorpdfstring{$T^{(3)}_n$}{T^3_n}}
Although only two derivatives come from the first $P_+$, because $P_+D^2P_+=0$
the rest of the derivatives must come from the next projection operator,
so the evaluation of $T_n^{(3)}$ is very similar to $T_n^{(2)}$. 
The result is
\begin{align*} 
	T_n^{(3)} &= \frac{(\Jb\J)^{n-2}}{16\Box^{n+1}}
		\frac{n(n+1)(n+2)}{60} 
		\Big(3\bar{a}a (n-1)(4n+2) + 15\bar{a}\J b (n-1) \\
		&\qquad + 5 a\Jb\bar{b} (3n+1) + 20\Jb\bar{b}\J b \Big)P_+\\
		&= \abar a \frac{(\Jb\J)^{n-2}}{16\Box^{n+1}}
		\frac{(n-2)(n-1)n(n+1)(n+2)}{30}P_+ + \text{surface terms} \ .
\end{align*}

\subsubsection*{Evaluation of \texorpdfstring{$T^{(4)}_n$}{T^4_n}}
One $D_\a$ and one $\Db_\da$ from the first projection operator hit fields
leaving
\begin{align*} 
	T_n^{(4)} &= \frac{1}{4\Box^{n+1}}\sum_{j=0}^{n-1}(j+1)^2
		(P_+\Jb\J)^{n-j-1}(\Jb\J)^{j}(\Db^\da\Jb)(D^a\J)\Db_\da D_\a P_+ \ .
\end{align*}
Since $\Db_\da D_\a P_+ = -2\rmi\pd_\ada P_+$, the next derivative can come
from any of the remaining projection operators. 
We sum over all possibilities and, 
after some work get the result 
\begin{align*} 
	T_n^{(4)} &= -a\frac{(\Jb\J)^{n-2}}{8\Box^{n+1}}
	\sum_{j=0}^{n-1}\sum_{k=0}^{n-j-2}
	(j+1)^2 (\Jb\bbar+(j+k+1)\abar)P_+ \\
	&= -a\frac{(\Jb\J)^{n-2}}{8\Box^{n+1}}\frac{(n-1)n(n+1)}{12}
	\big(n \Jb\bbar + \frac\abar{10}(8n^2-5n-2)\big)P_+ \\
	&= \abar a \frac{(\Jb\J)^{n-2}}{8\Box^{n+1}} 
	\frac{(n-2)(n-1)n(n+1)(2n-1)}{120}P_+ + \text{surface terms} \ .
\end{align*}

\subsubsection*{Evaluation of \texorpdfstring{$T^{(5)}_n$}{T^5_n}}
The evaluation of $T_n^{(5)}$ is similar to that of 
$T_n^{(4)}$, 
the final result is
\begin{align*}
	T_n^{(5)}
	&= -\frac{(\Jb\J)^{n-2}}{8\Box^{n+1}}\frac{n(n+1)(n+2)}{60}
	\big(\abar a (16 n^2 -13 n - 3) \\
	&\qquad	+ 5 \abar \J b (3n+1)
	+ 20 \Jb\bbar a (n-1) + 20 \Jb\bbar\J b \big)P_+ \\
	&= -\abar a \frac{(\Jb\J)^{n-2}}{8\Box^{n+1}} 
	\frac{(n-2)(n-1)n(n+1)(n+2)}{60}P_+ + \text{surface terms} \ .
\end{align*}

\subsubsection*{Total}
Combining all of the above, we find that 
\begin{align} \label{WZ-Tn-noIBP}
	T_n &= -\frac{(\Jb\J)^{n-2}}{16\Box^{n+1}}\frac{n(n+1)}{12}\Big(
	\abar a (8n^3+5n^2-11n-2) \\\non &\qquad
	+ 2\abar\J b (3n^2+5n+4) + 2a\Jb\bbar (5n^2+3n-8) 
	+ 2\Jb\bbar\J b (4n+5) \Big) P_+ \,,
\end{align}
which becomes remarkably simple after integration by parts
\begin{align} \label{WZ-Tn-IBP}
	T_n &= \abar a \frac{(\Jb\J)^{n-2}}{16\Box^{n+1}}
		\frac{n^2(n^2-1)}{12}P_+ 
		+ \text{surface terms}\ .
\end{align}

We can now calculate the 4-derivative correction to the EAFP
\begin{align*} 
	\G^{(1)}_{\text{4-deriv}} 
	&= \frac\rmi4 \Tr\sum_{n=1}^\infty\frac{-1}{n} \Big(T_n P_+ + \cc\Big) \, .
\end{align*}
Using the expression of $T_n$ after integration by parts \eqref{WZ-Tn-IBP} and 
moving to momentum space to diagonalise the trace, we have
\begin{align} \label{WZ-4D-DirectExp}
	\G^{(1)}_{\text{4-deriv}} 
	&= \frac1{(4\p)^2}\intz\frac{\abar a}{32}
		\int_0^\infty\rmd k\,\frac{k^{3}}{(k^2+\Jb\J)^4} \ .
\end{align}
Performing the final momentum integral yields a result of the form 
$\G^{(1)}_{\text{4-deriv}} = \intz \mathbb  F_{\text{4-deriv}}$,
where $\mathbb  F_{\text{4-deriv}}$ is given by 
\eqref{WZ-4Deriv-Zeta}
with 
\begin{align} 
\z=\frac1{384}~,
\label{384}
\end{align}
in agreement with the calculations of \cite{Pickering1996} and \cite{Pletnev1999}.

If, instead, we use \eqref{WZ-Tn-noIBP}, then, provided we 
integrate by parts before performing the momentum integral, 
we obtain the same result. 
However, if we leave the integration by parts until last, then 
each of the four terms in the momentum integral are IR divergent.
In which case, the momentum integrals can be performed if, e.g.,
we regularise with dimensional regularisation.
The result is
\begin{align} \label{WZ-4D-Direct-EpsExp}
	\G^{(1)}_{\text{4-deriv}} 
	&= \frac{(4\p\m^2)^\eps}{\G(2-\eps)(4\p)^2}\intz\frac{1/96}{(\Jb\J)^2}
	\Bigg( \frac{\abar a}{2}\Big(\frac1\eps-\log(\Jb\J)-\frac{13}{2}\Big)
	\!\!\\ \non
	&+(\abar\J b+\cc)\Big(\frac1\eps-\log(\Jb\J)+1\Big)
	-\frac{\Jb\bbar\J b}{2}\Big(\frac5\eps-5\log(\Jb\J)-9\Big) 
	\Bigg).
\end{align}
Integrating by parts, the $\frac1\eps$ and $\log$ terms cancel
and we once again recover the result 
\eqref{WZ-4Deriv-Zeta} 
with $\z$ given by \eqref{384}.

%%%%%%%%%%%%%%%%%%%%%%%%%%%%%%%%%%%%%%%%%%%%%%%%%%%%%%%%%%%%%%%%%%%%%%%%%%%%%%%%

\section{Auxiliary field potential via the heat kernel}
\setcounter{equation}{0}
\label{sec:AuxEffPot}

%%%%%%%%%%%%%%%%%%%%%%%%%%%%%%%%%%%%%%%%%%%%%%%%%%%%%%%%%%%%%%%%%%%%%%%%%%%%%%%%

In this section we examine the one-loop effective action 
of the WZ model starting with its 
expression in terms of the heat kernel 
$U_V^{(\J)}(s)$ studied in appendix \ref{sec:WZProp}. 
As shown in  \cite{Buchbinder1993, Buchbinder1994a}, the one-loop 
effective action may be represented in the form
\begin{align} \label{WZ-1Loop-HKEffAct}
	\G^{(1)} = \frac\rmi2\Tr\log(\frac\D\Box)
	= -\frac\rmi2\Tr\log(G_V^{(\J)})
	= -\frac\rmi2\int_0^\infty\frac{\rmd s}{s}\Tr U_V^{(\J)}(s) \,.
\end{align}	
In the effective potential limit, where $\pd_a\F=\pd_a\Fb=0$, 
it reduces to
\begin{align} \label{WZ-1Loop-HKEffAct-A}
	\G^{(1)} &= -\frac\rmi2\intz\int_0^\infty\frac{\rmd s}{s}
		\Big(A(s)+\tilde{A}(s)\Big)U(x,x'|s)\Big|_{x'\to x}\,,
\end{align}
which is the sum of the K\"ahler and auxiliary potentials
\begin{align}
		&= \intz \Big(K^{(1)} + \mathbb  F^{(1)} \Big)\ .
\end{align}
In the following subsections we evaluate the K\"ahler potential to check the above and to establish some notation. We then evaluate the four-derivative term in the auxiliary potential, first using the
integral introduced in \cite{Buchbinder1994a,Buchbinder1993}
\begin{align} \label{defn:BKY-dsJ}
	\dsJ(s) :=
	\frac2s \int_0^\infty \sin(p)\rme^{-p^2/s} \rmd p
	= \sqrt{\frac\p s}\rme^{-s/4}\mathrm{erfi}(\frac{\sqrt{s}}{2})\,,
\end{align}
and then using other methods to show that the term is actually conditionally convergent.
Finally we use the lessons learnt in the previous subsections to evaluate the full auxiliary potential.

%%%%%%%%%%%%%%%%%%%%%%%%%%%%%%%%%%%%%%%%%%%%%%%%%%%%%%%%%%%%%%%%%%%%%%%%%%%%%%%%

\subsection{K\"ahler potential}

In the K\"ahler approximation (see section \ref{ssec:WZProp:Kahler})
the diagonal of the heat kernel reduces to 
\begin{align} \label{eqn:Uv-Kahler}
\begin{aligned}
	U_V^{(\J)}(z,z|s) &= (\cos su - 1)(P_++P_-)\d^4(\q-\q')
		U(x,x'|s)\big|_{z'\to z} \\
		&= 2\Jb\J\frac{\cos su - 1}{u^2} U(x,x'|s)\big|_{x'\to x}
	\,,
\end{aligned}
\end{align}
where $u^2=\Jb\J\Box$ and 
$U(x,x'|s)$ is the dimensionally regularised bosonic heat kernel
defined in \eqref{DRegBosonicHK}.

The proper-time integral in \eqref{WZ-1Loop-HKEffAct} can then be evaluated
by first moving to momentum space. After Wick rotating and integrating out the angular parts
this leads to the following expression for the K\"ahler potential
\begin{align} \label{Kahler-skInt}
	K^{(1)} = \frac{-\mu^{2\eps}}{(4\p)^{2-\eps}}\frac{2}{\G(2-\eps)}
		\int_0^\infty \int_0^\infty k^{1-2\eps}
		\left(\cos\big(sk\left|\J\right|\big)-1\right)\rme^{-k^2s} \rmd k \rmd s \ .
\end{align}
The remaining integrals can be performed in either order to get the result 
\begin{align} \tag{\ref{eqn:1loopKahler}}
\begin{aligned} 
	K^{(1)} &= \frac{1}{2} \scJ(\Jb\J) 
	= \frac{\Jb\J}{2(4\p)^2}\left(\frac1\eps + 2 - \log\frac{\Jb\J}{\mub^2} + \ord\eps\right)\, ,
\end{aligned}
\end{align}
which matches \eqref{eqn:1loopKahler}.

Alternatively, we can follow \cite{Buchbinder1993, Buchbinder1994a} 
and swap dimensional regularisation for a proper-time cutoff $s_0 \to 0$.
Performing the momentum integral in \eqref{Kahler-skInt} now gives
\begin{align}
	K^{(1)} = \frac{\J\Jb}{2(4\p)^2}\int_{\rmi s_0}^\infty \frac{\rmd s}{s} \dsJ(s\J\Jb)\,,
\end{align}
where $\dsJ$ is defined in \eqref{defn:BKY-dsJ}.
This integral can be evaluated in terms of a hypergeometric function and expanded around $s_0 = 0$ to give
\begin{align}
	K^{(1)} = \frac{\J\Jb}{2(4\p)^2}\left( -\log(\rmi s_0 \m^2 \rme^\g) + 2 - \log\Big(\frac{\Jb\J}{\m^2}\Big) + \ord{s_0} \right)\,,
\end{align}
for some renormalisation scale $\m^2$.

%%%%%%%%%%%%%%%%%%%%%%%%%%%%%%%%%%%%%%%%%%%%%%%%%%%%%%%%%%%%%%%%%%%%%%%%%%%%%%%%

\subsection{Four-derivative term}

To find the coefficient of the leading term in the auxiliary potential, 
we need to evaluate \eqref{WZ-1Loop-HKEffAct-A} 
keeping only the four-derivative terms in the expression for $A(s)+\tilde{A}(s)$.
Expanding the results of section \ref{HK-results} 
gives
\begin{align} \label{WZ-AAt-4D}
	\hspace{-1.0em} 
	A(s)+\At(s){\Big|}_{\text{\smash{{4-deriv}}}}
	&= \frac{s\abar a}{512u^3}\Big(\big(7-\frac{10}{3}s^2u^2\big)\sin(su)
		+ su\big(s^2u^2-7\big)\cos(su)\Big) \non\\
	&+ \frac{s(\Jb\bbar a+\cc)}{64u^3}\Big(su\cos(su)
		- \big(1-\frac{s^2u^2}{3}\big)\sin(su)\Big) \\ \non
	&+ \frac{s \Jb\bbar\J b}{64u^3}\Big(\sin(su)-su\cos(su)\Big)   \ .
\end{align}

A general term in \eqref{WZ-AAt-4D} is of the form $\dsA=su^{-3}\dsa(su)$ 
and its contribution to the effective potential \eqref{WZ-1Loop-HKEffAct-A} is
\begin{align} \label{WZ-G1_dsA0}
	\G^{(1)}\big|_\dsA &= -\frac\rmi2\intz\int_0^\infty\frac{\rmd s}{s}
		\frac{s^4}{(su)^3}\dsa(su) U(x,x'|s)\Big|_{x'\to x} .
\end{align}
By using the $d$-dimensional momentum space representation for $U(x,x'|s)$, 
eq.\ \eqref{DRegBosonicHK},
then integrating out the angular parts of the momentum integral,
Wick rotating and rescaling the proper-time integral, we obtain
\begin{align} \label{WZ-G1_dsA}
	\!\G^{(1)}\big|_\dsA
	&= \frac{\m^{2\eps}}{\G(2-\eps)(4\p)^{d/2}}
	\int\!\!\frac{\rmd^8z}{(\Jb\J)^{2+\eps}}
	\int_0^\infty\frac{\rmd s}{s^{1-2\eps}}
		\int_0^\infty\frac{\rmd q}{q^{2\eps}}
		\dsa(q)\rme^{-\frac{q^2}{s}}\,, \!
\end{align}
where we've defined $q = s|k|\sqrt{\Jb\J}$. 
\pagebreak[2]

Removing the dimensional regularisation, 
it is now straightforward to use the definition \eqref{defn:BKY-dsJ}
in order to perform the momentum integral in \eqref{WZ-G1_dsA}
to write the four derivative contribution as 
\begin{align*} 
	\G^{(1)}_{\text{4-deriv}} &= \frac1{64(4\p)^2}\int\frac{\rmd^8z}{(\Jb\J)^2}
	\int_0^\infty\frac{\rmd s}{s}\Bigg( 
	\frac{s \Jb\bbar\J b}{4}\Big((s+2)\dsJ(s)-2\Big) \\
	&- \frac{s(\Jb\bbar a+\cc)}{24}\Big(\big(s^2+4s+12\big)\dsJ(s)-2(s+6)\Big)
	 \non\\
	&+ \frac{s\abar a}{384}\Big(\big(3s^3+2s^2+44s+168\big)\dsJ(s)
		-2\big(3s^2+8s+84\big)\Big)\Bigg)  .
\end{align*}
Each of the three terms in the above proper-time integral are IR divergent, 
but the divergences cancel when combined using integration by parts.
This gives a result of the form 
\begin{align} \tag{\ref{WZ-4Deriv-Zeta}}
	\mathbb  F_{\text{4-deriv}} 
	= \z\, \frac{(D^\a\F )(D_\a\F)(\Db_\da\Fb )(\Db^\da\Fb)}
		{(4\p)^2\left|m+\lambda\F\right|^4}\,,
\end{align}
where $\z=\z_{\BKY}$ is defined by the integral
\begin{align} \label{WZ-zetaBKS}
	\z_{\rm BKY} = 
	\frac1{1024}\int_0^\infty\!\!{\rmd s}\Bigg(1-\dsJ(s)
	+\frac{s}{2}(\dsJ(s)+4)-\frac{s^2}{4}(5\dsJ(s)+1)+\frac{s^3}{8}\dsJ(s) 
	\Bigg)\, .
\end{align}
Up to some  typographical errors, this matches equation (5.15) of 
\cite{Buchbinder1993, Buchbinder1994a}.
This integral can be evaluated to give the numerical result $\z_{\BKY} = -\frac1{64}$,
which clearly does not match the value of $\z=\frac1{384}$ found in the previous section.

Alternatively if we first integrate \eqref{WZ-AAt-4D} by parts to get the expression
\begin{align} \label{WZ-AAtIBP-4D}
	A(s)+\At(s){\Big|}_{\text{\smash{{4-deriv}}}}
	&\approx \frac{s\abar a}{1536 u^3}
	\Big( 3(1+2s^2u^2)\sin(su)-(3-s^2u^2)su\cos(su) \Big)~,
\end{align}
which holds up to surface terms, 
we can then evaluate \eqref{WZ-G1_dsA} without regularisation, 
as in the last paragraph, 
to find the four-derivative correction \eqref{WZ-4Deriv-Zeta} with
\begin{align*} 
	\z = 
	\frac1{1024}\int_0^\infty\!\!{\rmd s}\Bigg(\dsJ(s) - 1
	+\frac{s}{2}\Big(3\dsJ(s)+\frac83\Big)-\frac{s^2}{4}\Big(3\dsJ(s)+\frac13\Big)
	+\frac{s^3}{8}\frac13\dsJ(s) 
	\Bigg) ~. 	
\end{align*}
This result is different from \eqref{WZ-zetaBKS} 
and evaluates to the numerical value of $\frac1{192}$ which agrees with neither
of the previously found values for $\z$.

The problem lies in the fact that the unregularised ($\eps\to0$)
integrals are only conditionally convergent 
and not invariant under the rescaling required to obtain \eqref{WZ-G1_dsA}.  
If we don't perform the rescaling then it is convenient to try 
exchanging the order of the  proper-time and momentum integrals, 
as it leads to simpler intermediate expressions that are free from 
the error functions $\dsJ(s)$. 
However, when the order of the unregularised integrals is exchanged 
the result of the integration changes. 
This is a clear sign of conditional convergence.

If we keep the dimensional regularisation used in \eqref{WZ-G1_dsA} then we
consistently get the correction \eqref{WZ-4Deriv-Zeta} 
with $\z=\frac1{384}$. 
We demonstrate this with two possible order of operations.
First, we start with \eqref{WZ-AAtIBP-4D} and perform the proper-time
integral to get
\begin{align*} 
	\G^{(1)}_{\text{4-deriv}} 
	&= \frac{\m^{2\eps}}{\G(2-\eps)(4\p)^{d/2}}
	\intz\frac{\abar a}{32}
	\int_0^\infty{\!\!\rmd k}\,\frac{k^{3-2\eps}}{(k^2+\Jb\J)^4} \ .
\end{align*}
This momentum integral is clearly equivalent to \eqref{WZ-4D-DirectExp}
and converges for $-2<\eps<2$, 
so it does not need dimensional regularisation.
We recover the result \eqref{WZ-4Deriv-Zeta} with $\z=\frac1{384}$. 
However, if we start with \eqref{WZ-AAt-4D} and leave the integration by parts
until the very end, then we definitely need the dimensional regularisation.
Once again, for simplicity, performing the proper-time integral first, we find
\begin{align*} 
	\G^{(1)}_{\text{4-deriv}} 
	= \frac{(4\p\m^2)^\eps}{\G(2-\eps)(4\p)^2}
	\intz \frac{\Jb\J}{8} &\int_0^\infty \!\! \frac{\rmd k}{k^{1+2\eps}}
	\Big(\frac{\abar a}{12}\frac{5\Jb\J-4k^2}{(k^2+\Jb\J)^4} \\
	&- \frac{a\Jb\bbar+\cc}{3(k^2+\Jb\J)^3}
	+ \frac{b\bbar}{4(k^2+\Jb\J)^2}\Big)\ .
\end{align*}
The momentum integrals are IR divergent 
(i.e., in dimensional regularisation, they converge for $-2<\eps<0$)
and we get the $\eps$-expansion
\begin{align*} %\label{}
	\G^{(1)}_{\text{4-deriv}} 
	&= \frac{(4\p\m^2)^\eps}{\G(2-\eps)(4\p)^2}\intz \frac{1/96}{(\Jb\J)^2} 
	\Big(\frac{\abar a}{2}\Big(-\frac5{\eps} + 5\log(\Jb\J)-\frac{21}{2}\Big)\\
	&+ (a\Jb\bbar+\cc)\Big(\frac2\eps-2\log(\Jb\J)+3\Big)
	-\frac{3 \Jb \bbar \J b}{2}\Big(\frac1\eps-\log(\Jb\J)+1\Big) \Big) .
\end{align*}
This looks similar to \eqref{WZ-4D-Direct-EpsExp}, however the coefficients of the terms are different.
Nevertheless, integrating by parts yields the same
\eqref{WZ-4Deriv-Zeta} with $\z=\frac1{384}$.   

We note that in \cite{Buchbinder1994a,Buchbinder1993}, 
the action of $A(s)+\At(s)$ on $U(x,x'|s)$ 
was not evaluated by going to momentum space, 
but rather by series expansion and using \eqref{WZ-BosonicHK-DefRel}.
This leads to essentially identical results and problems to those discussed above. 
See the auxiliary {\it Mathematica} documents in \cite{TylerPhD2013}
for more details of this calculation and for other calculations using different regularisation schemes.

%%%%%%%%%%%%%%%%%%%%%%%%%%%%%%%%%%%%%%%%%%%%%%%%%%%%%%%%%%%%%%%%%%%%%%%%%%%%%%%%

\subsection{Full auxiliary field potential}
In the previous subsections, we have seen that the most robust and compact 
way to calculate the leading correction to the auxiliary potential
is to use the dimensionally regularised heat kernel, 
integrate by parts first, then perform the proper-time integral 
and finally the momentum space integral.
We will now follow this procedure 
to calculate the full one-loop EAFP.

The first step is to use integration by parts to get $A(s)+\At(s)$ into
a usable form.  Starting with the results \eqref{WZ-HKC-Solutions}
we find, after some work,
\begin{align*} 
	\J C(s) + \Jb \Ct(s)
	&\approx -2\rmi\Jb\J\frac{\sin(su)}{u} 
		-\rmi\frac{\abar a}{\bbar b}\Bigg(
		\Big(\frac{s^2u^2-1}{2u}-\frac{u}{\scF^2}\Big)\sin(su) \\
		&-\frac{3s}{2}\cos(su) 
		+\frac{\scG}{\scF}\Big(
		\frac{\cos(s\scF)\sin(s\scG)}{\scF}+\frac{\sin(s\scF)\cos(s\scG)}{\scG}
		\Big)\Bigg),
\end{align*}
which can then be integrated using \eqref{Aeqn} to get
\begin{align} \label{eqn:AplusAtIBP}
	A(s)+\At(s)
	&\approx 2\Jb\J\frac{\cos(su)-1}{u} \\\non
	&+\frac{\abar a}{\bbar b}\Bigg(
	\frac{s^2}{2}\Big(\cos(su)+\frac{\sin(su)}{su}\Big)
	+\frac{\cos(s\scF)\cos(s\scG)-\cos(su)}{\scF^2}\Bigg).
\end{align}
The first term is derivative free and corresponds 
to the K\"ahler approximation discussed above.
The second term contains all of the terms that generate the auxiliary potential,
starting with four derivative term \eqref{WZ-AAtIBP-4D}. 

Equation \eqref{eqn:AplusAtIBP} is an amazingly simple expression, 
considering the complexity of the results found in appendix \ref{sec:WZProp},
and is quite easily integrated to give the low-energy effective action.
The general structure is
\begin{align*} 
	\G^{(1)}
	&= \frac{\m^{2\eps}(4\p)^{-d/2}}{\G(2-\eps)}
	\intz\!\!\int_0^\infty\!\!\rmd k\, k^{3-2\eps}\!
	\int_0^\infty\!\frac{\rmd s}{s}\big(A(-\rmi s,u)+\At(-\rmi s,u)\big)
	\rme^{-k^2s}\ . 
\end{align*}
Performing the proper-time integral yields
\begin{align*} 
	\G^{(1)}
	&= \frac{\m^{2\eps}(4\p)^{-d/2}}{\G(2-\eps)}
	\intz\!\!\int_0^\infty\!\!\rmd k\, k^{3-2\eps}
	\Bigg[2\Jb\J\frac{\log(1+\Jb\J/k^2)}{2k^2\Jb\J} \\
	&+ \frac{\abar a}{\bbar b} \Bigg( 
	\frac{-1}{(k^2+\Jb\J)^2} 
	+\frac{2\log(\frac{\Jb\J}{k^2}+1)
	- \log\!\big(\big(\frac{\Jb\J}{k^2}+1\big)^2-\frac{4\scF^2}{k^2}\big)}{4\scF^2}
	\Bigg)\Bigg] \, .
\end{align*}
Factorising the final logarithm term,
the momentum integral can then be evaluated to get
\begin{align*} 
	\G^{(1)}
	&= \frac{\m^{2\eps}\G(\eps)}{(4\p)^{d/2}\G(2-\eps)}
	\intz\!\!
	\Bigg[\frac{\G(1-\eps)}{2(1-\eps)}(\Jb\J)^{1-\eps}
	- \frac{\abar a}{2\bbar b} \Bigg( 
	\frac{\G(2-\eps)}{(\Jb\J)^{\eps}} \\
	&+\frac{\G(1-\eps)}{(2-\eps)}\frac{2(\Jb\J)^{2-\eps}
		-(\Jb\J+2\scF)^{2-\eps}-(\Jb\J-2\scF)^{2-\eps}}{4\scF^2}
	\Bigg)\Bigg] \,.
\end{align*}
Expanding around $d=4$ and simplifying we get our result
\begin{align}
	\G^{(1)} = \intz \big( K^{(1)} + \mathbb  F^{(1)} \big)\,,
\end{align}
where the K\"ahler potential $K^{(1)}$ was given in \eqref{eqn:1loopKahler}
and the EAFP is
\begin{align} \label{WZ-fullAuxPot1}
\begin{aligned}
	(4\p)^{2}\mathbb  F^{(1)}
	= \frac{1}{4}\frac{\abar a}{\bbar b}\Bigg(&3 
	- \Big(1+\frac{16\Jb^2\J^2}{\bbar b}\Big)
	\log\Big(1-\frac{\bbar b}{16\Jb^2\J^2}\Big) \\
	&-\frac{16\Jb\J}{\sqrt{\bbar b}}\coth^{-1}
	\Big(\frac{4\Jb\J}{\sqrt{\bbar b}}\Big)\Bigg).
\end{aligned}
\end{align}
This has the series expansion
\begin{align} 
	(4\p)^{2}\mathbb  F^{(1)}
	&= \frac{\abar a}{4}\sum_{n=1}^{\infty}
	\frac{1}{n(n+1)(2n+1)}\frac{(\bbar b)^{n-1}}{(4\Jb\J)^{2n}} \\\non
	&= \frac{\abar a}{\Jb^2\J^2}\Big(
	\frac1{384}+\frac1{30720}\frac{\bbar b}{(\Jb\J)^{2}}
	+\frac1{1376256}\frac{(\bbar b)^2}{(\Jb\J)^{4}}+\dots\Big)\,,
\end{align}
where the natural expansion parameter is the dimensionless
\begin{equation} \label{defn:p2}
	p^2=\frac{\bbar b}{(4\Jb\J)^2}\, .
\end{equation}

Using integration by parts to remove $\abar a$ from the EAFP
essentially requires that we integrate $F^{(1)}$ with respect to $p$ twice.
This yields an expression with dilogarithms
\begin{align} \label{WZ-fullAuxPot2-series}
	(4\p)^{2}\mathbb  F^{(1)}
	&=  \frac{1}{4}\sum_{n=1}^{\infty}
	\frac{1}{n(n+1)(2n+1)(2n-1)^2}\frac{(\bbar b)^{n}}{(4\Jb\J)^{2n}} \\
	\label{WZ-fullAuxPot2}
	&= \frac{\Jb\J}{36}\Bigg(
	8+3p\Li_2(p)-3p\Li_2(-p) \nonumber\\
	&-\frac{1}{2p^2}
	\Big((p+1)\big(11p^{2}+7p+2\big)\log(p+1) + \big(p\to-p\big)\Big)
	\Bigg)\ .
\end{align}
This is reminiscent of \cite{DeWit1996,Pletnev1999,Banin2003c} where, 
for a $\cN=2$ SYM theory written in terms of $\cN=1$ superfields,
the one-loop K\"ahler potential was twice integrated to recover the
$\cN=2$ non-holomorphic potential. Their results were also expressed using
dilogarithms.

%%%%%%%%%%%%%%%%%%%%%%%%%%%%%%%%%%%%%%%%%%%%%%%%%%%%%%%%%%%%%%%%%%%%%%%%%%%%%%%%

\section{Comparisons to the component results}
\setcounter{equation}{0}
\label{sec:Components}

%%%%%%%%%%%%%%%%%%%%%%%%%%%%%%%%%%%%%%%%%%%%%%%%%%%%%%%%%%%%%%%%%%%%%%%%%%%%%%%%

Note that in the above two sections, the complicated and often poorly behaved expressions
simplified enormously after unifying the various terms through integration by parts.
In the component form of the effective potential with the background 
\begin{align}\tag{\ref{eq:spurion_background}}
	\F=\f+\q^2 f\,,  \quad \pd\f = \pd f= 0 \,,
 \end{align}
the functional forms are unique.\footnote{If 
instead of working with the effective potential, we are interested in the effective action, 
then the functional forms are once again not unique due to integration by parts identities in the spacetime integrals.}
This makes the component expressions a lot simpler to work with 
than their superfield 
counterparts.

The calculation of the full one-loop effective potential for the Wess-Zumino model 
has be performed many times before
\cite{Fujikawa1975, O'Raifeartaigh1976a,Huq1977,Amati1982,
Grisaru1983a,Miller1983a,Miller1983,Fogleman1983}
and does not need to be repeated here.
We will just quote the results, see \cite{TylerPhD2013} for more details.
The effective potential can be written as the dimensionally regularised momentum space integral
\begin{align}\label{WZCompEffPot-MtmInt}
	V^{(1)} = \frac{\rmi}{2}\int\!\!\frac{\rmd^dk}{(2\p)^d}\log
		\Big(1-\frac{|\l f|^2}{(k^2+| \jj  |^2)^2}\Big)\,,
\end{align}
where $\jj = \J|_{\q=\qb=0} = m+\l \f$. 
The momentum integral can be performed and yields
\begin{subequations}\label{WZCompEffPot}
\begin{align}
	(4\p)^2 V^{(1)} 
	\label{WZCompEffPot-Kahler}
	&= \frac{|\l f|^2}{2}\Big(-\frac1\eps+\log\frac{|\jj|^2}{\mub^2}+\ord\eps\Big)\\
	\label{WZCompEffPot-Auxiliary}
	&-|\l f|^2\left(\frac34 - \frac{|\jj|^2}{|\l f|}\tan^{-1}\frac{|\l f|}{|\jj|^2}
	 - \frac{1}{4}\Big(1+\frac{|\jj|^4}{|\l f|^2}\Big)\log\Big(1-\frac{|\l f|^2}{|\jj|^4}\Big)\right) .
\end{align}
\end{subequations}

Remembering that $\G^{(1)}=-\intx V^{(1)}$,
we see that the K\"ahler potential in superspace
projects to give the first line in \eqref{WZCompEffPot} 
through the relation
\begin{align}
	\intz K(\F,\Fb) = \intx \bar{f}f \pd_\f \pd_\fb K(\f,\fb) \ .
\end{align}
Equally as straightforward, the first line in \eqref{WZCompEffPot} 
can be lifted to superspace to give the K\"ahler potential by a simple double integral.

The derivative expansion of the EAFP is expressed using the dimensionless quantity $p^2$ defined in \eqref{defn:p2}.
It projects to the same ratio of component fields seen in \eqref{WZCompEffPot-Auxiliary}
\begin{align}
	p^2\big|_{\F=\f+\q^2 f} 
	= p_|^2 = \left|\frac{\l f}{\jj^2}\right|^2 \ .
\end{align}
Given the EAFP in the form $\Jb\J f(p^2)$, 
it can easily be projected to components using
\begin{align}
	\intz \Jb\J f(p^2) = \intx |\l f|^2 (1-p_|\pd_{p_|})^2 f(p_|^2) \ .
\end{align}
Equivalently, the component expression for the EAFP in the form $|\l f|^2 g(p_|^2)$ can be lifted 
to the superfield expression
$$\mathbb  F^{(1)}(\F,\Fb,a,\abar,b,\bbar) = \frac{\abar a}{(4\Jb\J)^2}\frac{g(p^2)}{p^2}\ .$$
Either way, we see that the EAFP given in \eqref{WZ-fullAuxPot1} and \eqref{WZCompEffPot-Auxiliary} are equivalent.

%%%%%%%%%%%%%%%%%%%%%%%%%%%%%%%%%%%%%%%%%%%%%%%%%%%%%%%%%%%%%%%%%%%%%%%%%%%%%%%%

\section{Conclusion and outlook}
\setcounter{equation}{0}
\label{sec:Concl}

%%%%%%%%%%%%%%%%%%%%%%%%%%%%%%%%%%%%%%%%%%%%%%%%%%%%%%%%%%%%%%%%%%%%%%%%%%%%%%%%

In this paper we have completed the calculation started in \cite{Buchbinder1994a,Buchbinder1993}
and used superfield techniques to compute 
the full one-loop supersymmetric effective potential for the WZ model. 
This includes both the effective K\"ahler potential and the previously 
unpublished result for the EAFP.
In the purely bosonic sector our results match the older component results for the effective potential of the WZ model. 

The supersymmetric effective potential 
contains more information than the ordinary effective potential of the WZ model. 
The point is that the superfield expressions \eqref{eqn:1loopKahler}
and \eqref{WZ-fullAuxPot1}  also generate two- and four-fermionic contributions
(generalised Yukawa couplings).
Of course, once the most general functional structure of the supersymmetric effective potential is known, one can read off the expressions for $K$ and $\mathbb  F$ 
from the component results. However, this functional structure became
available as a result of the superfield heat kernel calculation of 
\cite{Buchbinder1994a,Buchbinder1993}.

We have also compared different methods for calculating the leading term in the EAFP and accounted for the different results in the literature by noting that 
the calculation includes conditionally convergent integrals.
We observed that dimensional regularisation removes the conditional convergence 
and gives results that agree with the corresponding term in the component calculations.
It is interesting to observe that 
apparently finite terms in the effective action 
can result from conditionally convergent integrals in some calculation schemes.
This leads to possible ambiguities in calculations that can not 
be fixed by renormalisation conditions like those in nominally divergent terms.
One of the authors (SJT) has observed a similar problem occur in 
a two-loop effective action calculation in $\cN=2$ super-Yang-Mills.
How to identify when such cases can occur and how to fix the ambiguity is an open question.

The calculation of the leading term in the EAFP had possible IR divergences 
before the different structures were unified through integration by parts,
while the full EAFP does not appear to have such a problem and seems to not suffer any conditional convergence.
However, the expressions in the EAFP calculation are more unwieldy and this hinders their exploration. 
The momentum integrals in the EAFP are also difficult to perform as a whole 
and we used dimensional regularisation to handle the UV divergences that appeared when treating them separately.
The issue of any possible conditional convergence and ambiguity in the full EAFP is not completely closed.
Although, we should note that some of the component results were calculated using other regularisation schemes and they
yielded consistent results that match that which we presented in \eqref{WZ-fullAuxPot1}.

In calculating the EAFP at the one-loop level, 
we used the highly simplified structure of the heat kernel that occurs after integration by parts.
However, in higher loop calculations the full structure of the heat kernel is needed. 
This would make it an interesting challenge to try to carry out 
the superfield calculation of the EAFP to two loops.
Alternatively, the calculation could be performed with the background \eqref{eq:spurion_background}
and that result lifted to superspace using the results in the above section.
There are three published component calculations of the two-loop  effective potential of the WZ model 
\cite{Miller1984,Fogleman1983,Fogleman1984,SantosDos1989}.
However the results of \cite{Miller1984} were left as unevaluated Feynman integrals
and the results in \cite{Fogleman1983,Fogleman1984,SantosDos1989} contain terms that are less than quadratic in the auxiliary fields 
and thus can not come from the projection of a superfield action.
As for two-loop superfield calculations, both the effective K\"ahler potential and the effective chiral potential 
have been calculated many times, but the EAFP does not appear in the literature. 
It would be good to have a definitive result for it and the component effective
 potential at two loops.

In the massless case, 
as discussed in section 1, the one-loop EAFP of the massless WZ model 
can be chosen in  the form \eqref{1.15}
such that the functional $ \intz  \widetilde{\mathbb  F}^{(1)} $ is superconformal. 
This naturally leads us to consider a higher-derivative 
extension of the massless WZ model that is superconformal. 
It is described by an action of the form 
\bea
S[\F,\Fb] = \intz   \Fb\F \Big\{ 1
+ \bar \S \,\S \, \frak  H ( \S, \bar \S)\Big\}
&+&\l  \intc \F^3 + \bar \l \intac \Fb^3 ~,  \label{6.1} 
\eea 
where $\frak  H(w, \bar w)$ is a real analytic function and 
the chiral scalar 
\begin{align}
 \S &:=  \F^{-2}\bar D^2 \bar \F 
\end{align}
 is a (conformal) primary superfield of dimension zero.\footnote{The superfield $\X$
 defined by \eqref{1.13b} can be represented as $\X = \bar \F \F \bar \S \S$.}  
The massless WZ model possesses a ${\mathbb Z}_3$ symmetry generated by 
$\F \to \rme^{\frac{2}{3} \p \rmi } \F$. This symmetry remains intact for the 
higher-derivative extension of the WZ model given by  \eqref{6.1}, since $\S$ is invariant under the ${\mathbb Z}_3$ group. In the massless case, 
the one-loop effective action is invariant under U(1) phase transformations 
$\F \to \rme^{ \rmi \t} \F$. Requiring this symmetry gives 
$\frak  H (\S, \bar \S) = \hat{\frak  H} ( \S \bar \S)$. It would be interesting to see 
whether the U(1) invariant functional form of the EAFP, 
eq.\ \eqref{1.15},  survives at two loops. 

%%%%%%%%%%%%%%%%%%%%%%%%%%%%%%%%%
%%%%%%%%%%%%%%%%%%%%%%%%%%%%%%%%%

\appendix
%%%%%%%%%%%%%%%%%%%%%%%%%%%%%%%%%%%%%%%%%%%%%%%%%%%%%%%%%%%%%%%%%%%%%%%%%%%%%%%%

\section{Calculation of the heat kernel}
\setcounter{equation}{0}
\label{sec:WZProp}

%%%%%%%%%%%%%%%%%%%%%%%%%%%%%%%%%%%%%%%%%%%%%%%%%%%%%%%%%%%%%%%%%%%%%%%%%%%%%%%%

The superfield heat kernel \eqref{2.8} was computed 
in \cite{Buchbinder1994a} in the constant background case,  $\pd_a\J=\pd_a\Jb=0$.
The original derivation presented in \cite{Buchbinder1994a} 
contained some typographical errors. 
In this appendix we provide a corrected and simplified derivation
of the  heat kernel.
This is essential for our new results in section \ref{sec:AuxEffPot}.

\subsection{The differential equations for the heat kernel}
As we saw in section \ref{sec:Quant}, all propagators that occur in the 
Wess-Zumino model with arbitrary background chiral superfields $\J$ and $\Jb$ 
can be obtained as different chiral projections 
of the Green function of the operator 
\begin{align*} \tag{\ref{WZ-Green's Function}}
	\D
	=\Box-\frac{1}{4}(\J\Db^2+\Jb D^2)\ .
\end{align*}
The 
heat kernel of this operator obeys the differential equation (DE)
and initial condition
\begin{align} \label{eqn:WZHeatEqn}
	\left(\rmi\frac\rmd{\rmd s}+\D
		\right)U_V^{(\J)}(z,z'|s)&=0\,,\\
	U_V^{(\J)}(z,z'|0)
	&=\d^{4}(\q-\q')\d^4(x-x') \ .
\end{align}
If we assume that the background is constant over space-time, 
$\pd_a\J=\pd_a\Jb=0$, then the heat kernel factorises as 
\cite{Buchbinder1993, Buchbinder1994a}
\begin{align} \label{eqn:WZ:HeatKernFactorize} 
	U_V^{(\J)}(z,z'|s) = \O(s)U_V^{(0)}(z,z'|s)\,, \quad
	\O(s) 
	:=
	\rme^{-\frac{\rmi s}4\left(\Jb D^2+\J\Db^2\right)}\,,
\end{align}
where 
\( U_V^{(0)}(z,z'|s)
	=\d^4(\q-\q')U(x,x'|s)
\),
\begin{equation}
\label{WZ-BosonicHK-DefRel}
	\Box U(x,x'|s) = -\rmi\frac{\pd}{\pd s} U(x,x'|s)\,, 
\end{equation}
and
\begin{equation}
\begin{aligned}\label{4DBosonicHK}
	U(x,x'|s) 
	&= \exp(\rmi s\Box)\d^4(x-x')\\
	&= \int\frac{\rmd^4k}{(2\pi)^4}\rme^{-\rmi k^2 s+ \rmi k(x-x')} 
	= -\frac{\rmi}{(4\p s)^{2}}\rme^{\frac{\rmi}{4}(x-x')^2/s}\,,
\end{aligned} 
\end{equation}
is the 
free bosonic heat kernel. 
To find the full heat kernel \eqref{eqn:WZ:HeatKernFactorize}
we need only obtain an explicit form of the operator $\O(s)$.

The heat equation \eqref{eqn:WZHeatEqn} implies 
that the operator $\O(s)$ satisfies
\begin{align} \label{eqn:WZHeatEqn2}
	\rmi\frac\rmd{\rmd s}\O(s) = \frac14\O(s)\big(\J\Db^2+\Jb D^2)\,,
	\quad\O(0)=1 \ .
\end{align}
To solve this, following \cite{Buchbinder1993,Buchbinder1994a}, 
we expand the operator $\O(s)$ as
\begin{align}\label{eqn:decompOasABC}
	\O(s) &= \frac1{16}A(s)D^2\Db^2+\frac1{16}\tilde{A}(s)\Db^2D^2
	+\frac18B^\a(s)D_\a\Db^2+\frac18\tilde{B}_\da(s)\Db^\da D^2 \non\\
	&\qquad +\frac14C(s)D^2+\frac14\tilde{C}(s)\Db^2 + 1\ .
\end{align}
Note that only $A$ and $\tilde A$ can contribute to the 1-loop potential.
At this point, it is convenient to introduce some notation
\begin{align} \label{def:WZPropShortHands}
\begin{split}
	a=(D^\a\J)(D_\a\J)\,,\quad  b=(D^2\J)\,,\quad  
	\m=(D^\a\J)(\Db^\da\Jb) \pd_\ada \,, \\
	u^2=\Jb\J\Box\,,\quad
	\scF^2=\bbar b/64\,,\quad \scG^2=u^2+\scF^2 \,, \quad
	\b = \frac18\bem 0&\bbar\\ b& 0\eem .
\end{split}
\end{align}
We can move between the tilded and non-tilded symbols 
in \eqref{eqn:decompOasABC} by making the replacements 
$D^\a\leftrightarrow\Db_\da$, $\J\leftrightarrow\Jb$ which imply that
$a\leftrightarrow\abar$, $b\leftrightarrow\bbar$ and $\m\leftrightarrow-\m$.
Since we are using the convention that derivatives act on all terms to their
right unless bracketed, $\m$ is actually a
differential operator that obeys $\m^2=-\frac12\abar a \Box$.

With the above expansion, the heat equation \eqref{eqn:WZHeatEqn2} decomposes as 
\begin{subequations}
\begin{align}
 	\frac\rmd{\rmd s} \bem A\\\tilde A\eem
 	&= -\rmi	\bem\J&0\\0&\Jb\eem  \bem C\\\tilde C\eem ,\label{Aeqn}\\
 \left(\frac\rmd{\rmd s} + \bem 0&\J\pd^\daa\\\Jb\pd_\ada&0\eem \right)
 	\bem B^\a\\\tilde B_\da\eem
 	&= -\rmi\bem(D^\a\J)C\\(\Db_\da\Jb)\tilde{C}\eem , \label{Beqn} \\
 \left(\frac\rmd{\rmd s} + 2\rmi \b \right) \bem C\\\tilde{C} \eem
	 	+\rmi\, \bem\Jb (\Box A+1)\\\J(\Box\tilde A+1)\eem
  	&=-\frac{1}{2}\bem B^\a\pd_\ada(\Db^\da\Jb)\\
  		\tilde{B}_\da\pd^\daa(D_\a\J)\eem  , \label{Ceqn}
\end{align}
with $A(0)=\tilde{A}(0)=B^\a(0)=\tilde{B}_\da(0)=C(0)=\tilde{C}(0)=0$.
We can eliminate $A$ and $\tilde A$ from the equation for $C$ and $\tilde C$
by moving to the second order DE
\begin{align} \label{Ceqn2} 
 \left(\frac{\rmd^2}{\rmd s^2}+2\rmi\b 
 \frac\rmd{\rmd s}+\Jb\J\Box\right) \bem C\\\tilde{C}\eem
 =-\frac{1}{2}\frac\rmd{\rmd s}
 \bem B^\a\pd_\ada(\Db^\da\Jb)\\ \tilde{B}_\da\pd^\ada(D_\a\J)\eem  ,
\end{align}
\end{subequations}
where we need the initial ``velocity''
$\pd_s\big(C,\tilde{C}\big)\big|_{s=0}=-\rmi\big(\Jb,\J\big)$.

We solve the coupled equations (\ref{Beqn}, \ref{Ceqn2}) for $B$ and $C$
by expanding with respect to the Grassmann parameters $D_\a\J$ and $\Db_\da\Jb$, 
\begin{subequations}\begin{align}\label{cexpn}
\begin{split}
	C&=C_0+a C_{20}+\abar C_{02}+\m C_{11}+\abar a C_{22}\,,\\
	\Ct&=\Ct_0+\abar\Ct_{20}+a\Ct_{02}-\m\Ct_{11}+\abar a \Ct_{22}\,,
\end{split}\\
\label{bexpn}
\begin{split}
	B^\a&=(D^\a\J)(\hat{B}_0 + \abar\hat{B}_2)
	+(\Db_\da\Jb)\pd^\daa(\check{B}_0 + a \check{B}_2)\,,\\
  \Bt_\da&=(\Db_\da\Jb)(\hat\Bt_0 + a\hat\Bt_2)
	+(D^\a\J)\pd_\ada(\check\Bt_0 + \abar\check\Bt_2) \ .	
\end{split}
\end{align}
\end{subequations}
This expansion is used to find a system of ordinary second order 
differential equations for $B$ and $C$. 
For the rest of this subsection, 
we simply extract and list the DEs order by order in $D_\a\J$ and $\Db_\da\Jb$.
In the next subsection 
we note the common structure to the DEs and provide their solutions.

Keeping all terms independent of $D_\a\J$ and $\Db_\da\Jb$ 
in \eqref{Ceqn2} gives the homogeneous second order DE
\begin{gather}
	\left(\frac{\rmd^2}{\rmd s^2}+2\rmi \b\frac{\rmd}{\rmd s}+u^2\right)
	\bem C_0\\\Ct_0\eem =0 \,,
	\label{eqn:DE_C0} 
\intertext{with the initial conditions}
	\bem C_0\\\Ct_0\eem\Big|_{s=0}=0\,,	\quad
	\frac{\rmd}{\rmd s}\bem C_0\\\Ct_0\eem\Big|_{s=0}=-\rmi\JbJm \ .
	\label{eqn:IC_C0}
\end{gather}

Keeping only the first order terms in \eqref{Beqn} gives
\begin{align*}
\left(\frac\rmd{\rmd s}+\bem 0&\J\pd^\ada\\\Jb\pd_\ada&0\eem \right)
 \bem (D^\a\J)\hat{B}_0+(\Db_\da\Jb)\pd^\ada\check{B}_0
  \\(\Db_\da\Jb)\hat\Bt_0+(D^\a\J)\pd_\ada\check\Bt_0\eem
 =-\rmi\bem(D^\a\J)C_0\\(\Db_\da\Jb)\tilde{C}_0\eem  .
\end{align*}
Extracting the coefficients of $D_\a\J$ and $\Db_\da\Jb$ leads to two equations
that can be recombined to give the second order DE for $\check{B}_0$
\begin{align} 
	\left(\frac{\rmd^2}{\rmd s^2}+u^2\right)
	\bem\check{B}_0 \\ \check\Bt_0\eem
	&= \rmi \bem 0&\J\\ \Jb&0\eem\bem C_0\\\Ct_0\eem \,,
	\label{eqn:DE_Bc0}
\intertext{the solution of which immediately gives $\hat{B}_0$ 
			through the relation}
	\bem\hat{B}_0\\\hat\Bt_0\eem
	&=\frac{-1}{\Jb\J}\bem 0&\J\\\Jb&0\eem
	\frac\rmd{\rmd s}\bem\check{B}_0\\\check\Bt_0\eem .
	\label{eqn:DE_Bh0}
\end{align}

Keeping only the second order terms in \eqref{Ceqn2} yields
the two second order DEs
\begin{align}
	\left(\frac{\rmd^2}{\rmd s^2}+2\rmi\b\frac{\rmd}{\rmd s}+u^2\right)
	 \bem a C_{20}+\abar C_{02}\\\abar\Ct_{20}+a\Ct_{02}\eem 
	 &=\frac\Box2\frac\rmd{\rmd s} 
	 	\bem \abar \check B_0 \\ a \check\Bt_0\eem , 
	 \label{eqn:DE_C2}\\
	\left(\frac{\rmd^2}{\rmd s^2}+2\rmi\b\frac{\rmd}{\rmd s}+u^2\right)
		 \bem \phantom{-}C_{11} \\ -\Ct_{11}\eem 
	 &=-\frac12\frac{\rmd}{\rmd s} 
		 \bem\phantom{-}\hat B_0 \\ -\hat\Bt_0 \eem .
	\label{eqn:DE_C11}
\end{align}
Although we could separate the DEs for $C_{20}$ and $C_{02}$, it is simpler
(more symmetric) to solve for them simultaneously.

Keeping only the third order terms in \eqref{Beqn} gives
\begin{align}\begin{split}
	\left(\frac\rmd{\rmd s}+\bem 0&\J\pd^\daa\\\Jb\pd_\ada&0\eem \right)
 	\bem \abar (D^\a\J)\hat{B}_2+a(\Db_\db\Jb)\pd^\dba\check{B}_2\\
  		a(\Db_\da\Jb)\hat\Bt_2+\abar (D^\b\J)\pd_\bda\check\Bt_2\eem \\
    =-\rmi\bem(D^\a\J)(\abar C_{02}+\m C_{11})\\
  			(\Db_\da\Jb)(a\Ct_{02}-\m\Ct_{11})\eem .
\end{split}\end{align}
Using 	\( (D^\a\J)\m=\frac{1}{2} a(\Db_\da\Jb)\pd^\daa \)
and		\( (\Db_\da\Jb)\m=-\frac{1}{2}\abar(D^\a\J)\pd_\ada \)
we can split the above to get the second order DE for $\check{B}_2$
\begin{align}
\left(\frac{\rmd^2}{\rmd s^2}+u^2\right)\bem\check{B}_2\\\check\Bt_2\eem
	&=\rmi\bem 0&\J\\\Jb&0\eem\bem C_{02}\\\Ct_{02}\eem
	 -\frac\rmi2\frac{\rmd}{\rmd s}\bem C_{11}\\\Ct_{11}\eem \,,
	\label{eqn:DE_Bc2}
\intertext{the solution of which immediately gives $\hat{B}_2$ 
			through the relation}
	\bem\hat{B}_2\\\hat\Bt_2\eem&=\frac{-1}{\Jb\J}\bem 0&\J\\\Jb&0\eem\left(
	\frac\rmd{\rmd s}\bem\check{B}_2\\\check\Bt_2\eem 
	+\frac\rmi2\bem C_{11}\\\Ct_{11}\eem\right)  .
	\label{eqn:DE_Bh2}
\end{align}

The final DE is easily read from the highest order terms in \eqref{Ceqn2},
\begin{align}\label{eqn:DE_C22}
 	\left(\frac{\rmd^2}{\rmd s^2}+2i\b\frac{\rmd}{\rmd s}+u^2\right)
	\bem C_{22}\\\Ct_{22}\eem
	=\frac\Box2\frac\rmd{\rmd s}\bem\check{B}_2\\\check\Bt_2\eem~.
\end{align}

%%%%%%%%%%%%%%%%%%%%%%%%%%%%%%%%%%%%%%%%%%%%%%%%%%%%%%%%%%%%%%%%%%%%%%%%%%%%%%%%

\subsection{Results for the heat kernel}\label{HK-results}
The differential equations \eqref{eqn:DE_C0} for $(C_0,\Ct_0)$ 
are the only ones that are both homogeneous 
and have non-vanishing initial conditions. 
So, their integration is straightforward, 
with the results given in \eqref{WZ-HKC0-Solution}.
The DEs that need to be solved to find the terms of higher order 
in $D_a\J$ and  $\Db_\da\Jb$
all second order, inhomogeneous DEs with vanishing initial conditions.
That is, they are all of the form
\begin{subequations}\begin{align}
	\left(\frac{\rmd^2}{\rmd s^2}+2\rmi\b\frac{\rmd}{\rmd s}
		+u^2\right)\c_C(s) &= v_C(s)\,,
	\qquad \c_C(0)=\dot\c_C(0)=0 \,, \\
	\left(\frac{\rmd^2}{\rmd s^2}+u^2\right)\c_B(s) &= v_B(s) \,,
	\qquad \c_B(0)=\dot\c_B(0)=0 \,,
\end{align}\end{subequations}
where the $\c_{B,C}$ are component 2-vectors of $B$ or $C$ respectively 
and the inhomogeneous terms $v_{B,C}$ depend on 
the solutions to lower order components.
Using variation of parameters on the general solutions to the 
associated homogeneous differential equations yields \vspace{-1.5em}
\begin{subequations}\begin{align}
	\c_C(s)&=\rme^{\rmi s\b(\scG/\scF-1)}
	\int_0^s\!\!\rmd{t}\,\rme^{-2\rmi t\b\scG/\scF}
	\int_0^t\!\!\rmd{\t}\,\rme^{\rmi\t\b(\scG/\scF+1)}v_C(\t) \,, \\
	\c_B(s)&=\rme^{\rmi su}\int_0^s\!\!\rmd{t}\,\rme^{-2\rmi tu}
	\int_0^t\!\!\rmd{\t}\,\rme^{\rmi\t u}v_B(\t) \ .
\end{align}\end{subequations}
The following solutions have all been found by hand 
and checked that they satisfy the original DEs 
and boundary conditions using \emph{Mathematica}. 
The solutions for the components of $\big(C(s),\Ct(s)\big)$ are
\begin{subequations} \label{WZ-HKC-Solutions}
\begin{align}
\label{WZ-HKC0-Solution}
\bem C_0\\\Ct_0\eem 
	&= -\rmi\frac{\sin(s\scG)}{\scG}\rme^{-\rmi s\b}\JbJm 
~	,\\
\bem C_{11}\\\Ct_{11}\eem 
	&= \frac{s}{8\scF^2}\left(\frac{\sin(su)}{su}
		-\frac{\sin(s\scG)}{s\scG}\cos(s\scF)\right)\JbJm ,
 \allowdisplaybreaks\\
\bem C_{20}\\\Ct_{20}\eem 
 	&= \frac{\Box\b}{8\scF^2}\bem 0&\J\\\Jb&0\eem \!\! \Bigg[\
		\frac{\rmi s}{2u^2}\left(\frac{\sin(s\scF)}{s\scF} \cos(s\scG)
		-\cos(s\scF)\frac{\sin(s\scG)}{s\scG}\right) \non \\  \Bigg.
		& +\frac{\b}{u^2}\left(
		\frac{\cos(su)}{\scF^2}-\frac{\sin(s\scF)\sin(s\scG)}{\scF\scG}-
		\frac{\cos(s\scF)\cos(s\scG)}{\scF^2}\right) \\ \non
		& -\frac{\rmi s}{2\scG^2}\left(\cos(s\scG)
		-\frac{\sin(s\scG)}{s\scG}\right)
		\rme^{-\rmi s\b} \Bigg]\JbJm ,  
 \allowdisplaybreaks\\
\bem C_{02}\\\Ct_{02}\eem 
	&= \frac{\Box}{16}\bem 0&\J\\\Jb&0\eem\frac{-\rmi\b}{\scF^2}\Bigg[
		\frac{\sin(s\scG)\cos(s\scF)}{u^2\scG}
		-\frac{\sin(s\scF)\cos(s\scG)}{u^2\scF} \\\non
		&+\frac{s\rme^{-\rmi s\b}}{\scG^2}
		\left(\cos(s\scG)-\frac{\sin(s\scG)}{s\scG}\right)\!\Bigg]\!\JbJm ,
 \allowdisplaybreaks\\
\bem C_{22}\\\Ct_{22}\eem
	&= \frac{-\rmi\Box}{128\scF^2}\Bigg[
		\frac{\rmi s^2\b}{\scF^2}
		  \left(\frac{\sin(s\scF)}{s\scF}\frac{\sin(s\scG)}{s\scG}
		  -\frac{\sin(su)}{su}\right)
		+ s\,\rme^{-\rmi s\b} \times \\ \non
		&\times\!\left(\!\frac{\sin(s\scG)}{s\scG}\left(
			\frac{1+\rmi s\b}{\scF^2}
			-\frac{3\scF^2+(1+s^2u^2)\scG^2}{2\scG^4} \right) 
			+ \cos(s\scG)\frac{3\scF^2+\scG^2}{2\scG^4}\right)\\ \non
		&-\frac{s}{u^2}\left(\frac{u^2}{\scF^2}\frac{\sin(su)}{s u}
			+\frac{\sin(s\scF)}{s\scF}\cos(s\scG)
			-\cos(s\scF)\frac{\sin(s\scG)}{s\scG}\right) \Bigg]\JbJm ,
\end{align}\end{subequations}
and the solutions for the components of $\big(B^\a(s),\Bt_\da(s)\big)$ are
\begin{subequations} \label{WZ-HKB-Solutions}
\begin{align}
\bem \check{B}_0\\\check{\Bt}_0\eem
	&= \frac{\rmi s}{2u^2}\bem 0&\J\\\Jb&0\eem\left(
		\frac{\b}{\scF}\frac{\cos(su)}{s\scF}
 		-\rme^{-\rmi s\b}\left(\rmi\frac{\sin(s\scG)}{s\scG}
		+\frac{\b}{\scF}\frac{\cos(s\scG)}{s\scF}\right)\!\right)\!\JbJm ,\\ 
\bem \hat{B}_0\\\hat{\Bt}_0\eem
	&= \frac{\rmi s\b}{2\scF^2}\left(\frac{\sin(su)}{su}
		-\frac{\sin{s\scG}}{s\scG} \rme^{-\rmi s\b}\right)\JbJm ,
 \allowdisplaybreaks\\
\bem \check{B}_2\\\check{\Bt}_2\eem
	&= \frac{-\rmi s^2}{32\scF^2}\Bigg[
		\frac{2\rmi\b}{su^2}\Big(
			\frac{\scG^2}{\scF^2}\frac{\sin(s\scG)}{s\scG}\cos(s\scF)
			-\cos(s\scG)\frac{\sin(s\scF)}{s\scF}\Big)		\\\non
	&+ \frac{\sin(su)}{su}\Big(1-\frac{2\rmi\b}{s\scF^2}\Big)
		-\rme^{-\rmi s\b}\Bigg(\!\!
			\Big(1+\frac{\rmi\b}{s\scG^2}\Big)\frac{\sin(s\scG)}{s\scG}
			-\frac{\rmi\b}{\scG}\frac{\cos(s\scG)}{s\scG}\Bigg)\!
		\Bigg]\!\JbJm \!, 
 \allowdisplaybreaks\\
\bem \hat{B}_2\\\hat{\Bt}_2\eem
	&= \frac{-\rmi}{\Jb\J}\frac{1}{16\scF^2}\bem 0&\J\\\Jb&0\eem\Bigg[
		\frac{\sin(su)}{2u}-\cos(su)\frac{s\scF^2-2\rmi\b}{2\scF^2} \\ \non
		&\quad -\left(	\frac{\sin(s\scG)}\scG
		+\frac{\rmi\b}{\scF^2}\cos(s\scG)\right)\cos(s\scF) \\ \non
		&\quad +\frac{1}{2\scG^2}\left(u^2s\cos(s\scG)+(\scF^2+\scG^2)
		\frac{\sin(s\scG)}\scG \right) \rme^{-\rmi s\b}\Bigg]\JbJm \ .
\end{align}
\end{subequations}
The solution for $\big(A, \At \big)$ is just 
a term-by-term integration \eqref{Aeqn} of the expression for $\big(C, \Ct \big)$ 
given in \eqref{WZ-HKC-Solutions} above. 

From the above results, it is easily checked that 
the solutions satisfy the initial condition $\O(0)=1$.
They also satisfy the initial velocity condition
$\O'(0)=-\frac\rmi4(\Jb D^2 + \J\Db^2)$,
which implies that only $\big(C_0,\Ct_0\big)$ has a non-vanishing first derivative at $s=0$.

The above results for the coefficients in the expansion of 
$\O(s) = \exp\left(-\frac{\rmi s}4\left(\Jb D^2+\J\Db^2\right)\right)$ 
combined with the factorisation 
\begin{align} 
\nonumber
	U_V^{(\J)}(z,z'|s) = \O(s)U_V^{(0)}(z,z'|s)\,, 
	\quad U_V^{(0)}(z,z'|s) = \d^4(\q-\q')U(x,x'|s)\,,
\end{align}
give the full solution for the heat kernel of the Wess-Zumino model in four dimensions
in terms of the bosonic heat kernel \eqref{4DBosonicHK}.
The dimensionally reduced heat kernel is obtained by simply replacing the 
momentum integral in \eqref{4DBosonicHK} with its dimensionally reduced counterpart 
\begin{equation}
\begin{aligned}\label{DRegBosonicHK}
	U(x,x'|s) 
	&= \exp(\rmi s\Box)\d^d(x-x')\\
	&= \m^{2\eps}\int\frac{\rmd^dk}{(2\pi)^d}\rme^{-\rmi k^2 s+ \rmi k(x-x')} 
	= \frac{\rmi\m^{2\eps}}{(4\p\rmi s)^{d/2}}\rme^{\frac{\rmi}{4}(x-x')^2/s}\  .
\end{aligned} 
\end{equation}

%%%%%%%%%%%%%%%%%%%%%%%%%%%%%%%%%%%%%%%%%%%%%%%%%%%%%%%%%%%%%%%%%%%%%%%%%%%%%%%%

\subsection{K\"ahler approximation}\label{ssec:WZProp:Kahler}
As a check on the general results above, 
we examine the limit of the heat kernel that is appropriate for computing 
the corrections to the K\"ahler potential.
That is, we enforce the condition $\J=\const$ by taking the limits as 
$a$, $b$, and $\m$ go to $0$, which implies that 
$u^2=\cG^2=\Jb\J\Box$ and $A=\tilde{A}=\Jb\J(\cos(su)-1)/{u^2}$.
So, the expansion of $\O(s)$ reduces to
\begin{align} \label{WZ:O(s)-in-Kahler-approx}
	\O(s) 
		  = 1 - \frac{\rmi}{4}\frac{\sin su}{u}\left(\Jb D^2 + \J\Db^2\right)
		  	  + \frac{\Jb\J}{16}\frac{\cos su - 1}{u^2}\acom{D^2}{\Db^2} \,,
\end{align}
which matches that presented in \cite{Buchbinder1993,Buchbinder1994a,BK}.

This result can also be derived directly from \eqref{eqn:WZHeatEqn2}.
When $\J=\const$, we can take a second proper-time derivative
to find the inhomogeneous harmonic oscillator equation%
\footnote{In deriving this differential equation we used 
the superspace projection operators given in \eqref{defn:SuSyProjOps}.
}
\begin{align*} 
	\O''(s) = -\frac1{16}\O(s)(\Jb D^2 + \J\Db^2)^2
		= -u^2\O(s)(P_++P_-)
		= -u^2\O(s) + u^2 P_0\,,
\end{align*}
with the initial conditions $\O(0)=1$, $\O'(0)=-\frac\rmi4(\Jb D^2 + \J\Db^2)$.
This is easily solved to give \eqref{WZ:O(s)-in-Kahler-approx}.

%%%%%%%%%%%%%%%%%%%%%%%%%%%%%%%%%%%%%%%%%%%%%%%%%%%%%%%%%%%%%%%%%%%%%%%%%%%%%%%%
%%%%%%%%%%%%%%%%%%%%%%%%   Bibliography   %%%%%%%%%%%%%%%%%%%%%%%%%%%%%%%%%%%%%%
%%%%%%%%%%%%%%%%%%%%%%%%%%%%%%%%%%%%%%%%%%%%%%%%%%%%%%%%%%%%%%%%%%%%%%%%%%%%%%%%

\begin{footnotesize}
%\providecommand{\href}[2]{#2}
%\bibliographystyle{sjtphys} %unsrt
%\pdfbookmark[1]{References}{references_bookmark}
%\bibliography{abbrev} 

% bibliography copied from generated BBL file
\providecommand{\href}[2]{#2}\begingroup\raggedright
\endgroup

\end{footnotesize}

\end{document}